\documentclass[useAMS,usenatbib]{mn2e}
\usepackage{mn2ejour}
\usepackage{graphicx}
\title[Planet formation on submillimetre images]{Possible planet-forming regions on submillimetre images}

\author[Zs. Reg\'aly, A. Juh\'asz, Zs. S\'andor and C. P. Dullemond]{Zs. Reg\'aly$^{1}$\thanks{E-mail:regaly@konkoly.hu}, A. Juh\'asz$^{2}$, Zs. S\'andor$^{3}$ and C. P. Dullemond$^4$\\
$^{1}$Konkoly Observatory of the Hungarian Academy of Sciences, P.O. Box 67, H-1525 Budapest, Hungary; regaly@konkoly.hu\\
$^2$Leiden Observatory, Leiden University, P.O. Box 9513, NL-2300 RA Leiden, The Netherlands\\
$^3$Max Planck Research Group, Max-Planck-Institut f\"ur Astronomie, K\"onigstuhl 17, D-69117 Heidelberg, Germany\\
$^4$Institut f\"ur Theoretische Astrophysik, Universit\"at Heidelberg, 69120 Heidelberg, Germany}

\begin{document}

\date{Accepted 2011 September 14.  Received 2011 September 14; in original form 2011 April 25}

\pagerange{\pageref{firstpage}--\pageref{lastpage}} \pubyear{2011}

\maketitle

\label{firstpage}

\begin{abstract}
Submillimetre images of transition discs are expected to reflect the distribution of the optically thin dust. Former observation of three transition discs LkH$\alpha$\,330, SR\,21N, and HD\,1353444B at submillimetre wavelengths revealed images which cannot be modelled by a simple axisymmetric disc. We show that a large-scale anticyclonic vortex that develops where the viscosity has a large gradient (e.g., at the edge of the disc dead zone), might be accountable for these large-scale asymmetries. We modelled the long-term evolution of vortices being triggered by the Rossby wave instability. We found that a horseshoe-shaped (azimuthal wavenumber $m=1$) large-scale vortex forms by coalescing of smaller vortices within $5\times10^4\,\rm yr$, and can survive on the disc life-time ($\sim5\times10^6\,\rm yr$), depending on the magnitude of global viscosity and the thickness of the viscosity gradient. The two-dimensional grid-based global disc simulations {with local isothermal approximation and compressible-gas model} have been done by the GPU version of hydrodynamic code {\small FARGO} ({\small GFARGO}). To calculate the dust continuum image at submillimetre wavelengths, we combined our hydrodynamical results with a 3D radiative transfer code. By the striking similarities of the calculated and observed submillimetre images, we suggest that the three transition discs can be modelled by a disc possessing a large-scale vortex formed near the disc dead zone edge. Since the larger dust grains (larger than mm in size) are collected in these vortices, the non-axisymmetric submillimetre images of the above transition discs might be interpreted as active planet and planetesimal forming regions situated far ($\ga 50\,\rm AU$) from the central stars.
\end{abstract}

\begin{keywords}
accretion, accretion discs --- protoplanetary discs --- hydrodynamics --- instabilities --- methods: numerical)
\end{keywords}

\section{Introduction}

Recently \citet{Brownetal2009} presented submillimetre images of three transition discs LkH$\alpha$\,330, SR\,21N, HD\,135344B. Using a 2D radiative transfer model \citep{DullemondDominik2004}, they showed that the observed inner discs are cleared from dust as the optical through millimetre-wave spectral energy distribution (SED) can be fitted with a disc having an inner cavity. The non-axisymmetric intensity distributions (see fig. 1 of \citealt{Brownetal2009}), however, require further assumptions such as the presence of an embedded planet at large distances from the central star ($\ga50$\,AU) that sufficiently perturbs the disc dust distribution (see, e.g., \citet{Wyattetal1999}). A massive enough planet opens a gap, and the disc's material will be piled up just outside of the giant planet's orbit \citep{deValBorroetal2007}. Although the giant planet induced non-axisymmetric density distribution hypothesis is challenging, we propose in this Paper that the observed asymmetries are due to large observable vortices formed in these discs during their early evolutionary stages.

However, the current most accepted scenarios for giant planet formation, the core-accretion \citep{BodenheimerPollack1986,Pollacketal1996} and the gravitational instability \citep{Boss2001,Boss2006a} face problems forming and keeping giant planets at such a large distances. Within the core-accretion scenario, the optimal places for planet formation is the vicinity of the snow line where the condensation of volatiles into ices increases the surface density of solids by a large factor, increasing the isolation mass above the critical core mass. The location of the snow line determined by the opacity of the dust grains and the mass accretion rate is recently found to be $\la 16\,\rm AU$ \citep{Minetal2011}. For significantly larger distances, the growth rate of the planetary core is strongly limited. On the other hand, forming giant planets on wide orbits via gravitational instability does not face many problems (assuming rapid cooling of the gas), except that it is not yet clear how to keep them there. \citet{Dodson-Robinsonetal2009} investigated three possible mechanisms -- core-accretion (with or without migration), scattering from the inner disc, or gravitational instability -- which might result in massive planets at wide ($\ga35$\,AU) orbits discovered by high-contrast direct imaging observations of Fomalhaut \citep{Kalasetal2008mn}, HR\,8799 \citep{Maroisetal2008} or CD-352722 \citep{Wahhajetal2011mn} for instance. Dodson-Robinson et al. found that the only possible scenario to form stable wide-orbit planets is the gravitational instability, but planets formed this way were recently shown (by \citealp{Baruteauetal2011}) to migrate quite rapidly inwards -- puts in question its feasibility. Note that \citep{Cridaetal2009} have shown that runaway outward migration by planets in mean motion resonance carving a common gap in the protoplanetary disc is an other possible mechanism to place giant planets on wide orbits, if the inner one is significantly more massive than the outer one.


In recent years the formation of vortices in protoplanetary discs have been investigated thoroughly, since they may play an important role in the formation of planets in the core-accretion scenario. Observation of such vortices would have a great importance providing further informations and support for planetary formation theories (see, e.g., \citet{KlahrHenning1997,BargeSommeria1995}).

The classical approach of accretion discs \citep{ShakuraSunyaev1973} postulates a simple scaling relation between the gas viscosity and the level of angular momentum transport parametrized by $\alpha$. An accretion disc coupled dynamically to a weak magnetic field is subject to the so-called magnetorotational instability (MRI) resulting in outward angular momentum transport, which can be described by an effective $\alpha$-viscosity \citep{BalbusHawley1991}. The disc becomes MRI-active, if the ionisation state of the gas is sufficient. Investigating the ionisation processes in MRI-active discs, \citet{Gammie1996} proposed the existence of a radially extended region in the disc midplane, where the accretion is strongly reduced. Near the disc inner edge, the midplane gas is collisionally ionised, owing to the high temperatures \citep{PneumanMitchell1965,Umebayashi1983}. Further out, the temperature is not high enough to collisionally ionise the gas, while the gas is yet dense enough to shield the interstellar cosmic-rays or diluted stellar X-rays. The magnetic field of the ionised disc corona might even deflect the cosmic rays. As a result, only a thin surface layer of the disc is ionised at this region, while at even larger distances, the cosmic rays can penetrate the tenuous disc which might ionise the disc midplane again \citep{Glassgoldetal1997,IgeaGlassgold1999}. As a consequence, the turbulent viscosity is low in the poorly ionised region of the disc midplane, resulting in the formation of the so called {\it dead zone}. Within this region the viscous accretion is taking place mostly in a narrow upper layer of the disc, while it is  stalled in the disc's midplane.

\citet{VarniereTagger2006} and \citet{Terquem2008} demonstrated that density and pressure enhancements (`bumps') appear at the boundaries of a \emph{dead zone} because of the sharp jump in the viscosity. These pressure bumps are found to be unstable to the Rossby wave instability (RWI) \citep{Lovelaceetal1999,Lietal2000}, leading to the formation of a large-scale anticyclonic vortices.

\citet{KretkeLin2007} have already shown that pressure-gradient inversion formed at the inner edge of the dead zone or at the snow line causes the accumulation of dust. The existence of large-scale vortices can also accelerate the formation of planetesimals and planetary embryos in the core-accretion paradigm. \citet{BargeSommeria1995} and \citet{KlahrHenning1997} have already shown that particles tend to get trapped in anticyclonic vortices. At the radial location of the vortex the type-I migration is halted due to radial pressure-gradient inversion, thus not only dust is retained here but larger bodies too. Particles of about centimetre to metre in size subsequently drift into the pressure maxima, i.e., to the vortices \citep{Lyraetal2009,Katoetal2010,Dzyurkevichetal2010} and form gravitationally bound clumps of solids, which can coalesce forming embryos between the masses of the Moon and Mars. The swarm of these embryos evolves further by mutual N-body collisions forming massive cores ($\sim 10 M_{\oplus}$) of giant planets in $\leq 5\times10^5\,\rm yr$ \citep{Sandoretal2011}. Thus, large-scale anticyclonic vortices formed in discs are extremely important for the planetary formation helping to overcome the meter-size barrier problem \citep{BlumWurm2008}, and the time-scale problem of oligarchic growth. From this perspective these vortices can be regarded as `planetary factories'.

The micron sized dust is dynamically coupled to the gas as it is indirectly supported against the settling by gas pressure via the gas drag, thus their spatial distribution is expected to be the same. 
Assuming that the three transition discs presented in \citet{Brownetal2009} contain significant amount of gas -- supported by H\,$\alpha$ \citep{Fernandezetal1995}, [OI] \citep{vanderPlasetal2008} and fundamental band CO emissions \citep{Pontoppidanetal2008,Salyketal2009} --, these vortices can be imaged in submillimetre wavelengths as overdense regions in the dust. The inner holes in the submillimetre images of the three discs may indicate that dust is either consumed here by a planet forming process or by coagulation to larger sized grains, which are drifted rapidly into the star. Note that coagulation without fragmentation would make the vortex also invisible as the dust opacity is decreasing with grain size. If the dust trapped inside giant vortices \citep{InabaBarge2006} is accountable for the observed features, fragmentation should be happening inside them, in order to replenish the population of small dust grains.

Inspired by these ideas, we explore, whether a large-scale, long-lived (on $10^6\,\rm yr$ time-scale) non-axisymmetric gas-dust perturbation formed at the outer edge of the disc dead zone could explain the non-axisymmetric submillimetre images. To calculate the density perturbation, we run long-term numerical hydrodynamical simulations on a GPU aided computer. In \S\,2, we present our hydrodynamical disc model, and a study on the robustness of the formation of large-scale vortices. Submillimetre dust continuum images calculated by a 3D radiative transfer using our hydrodynamical results are presented in \S\,3. The Paper closes (\S\,4) with the discussion and concluding remarks, along with a short discussion on the reasonability of our findings.

\section{Vortex formation}

\begin{figure}
	\centering
	\includegraphics[width=\columnwidth]{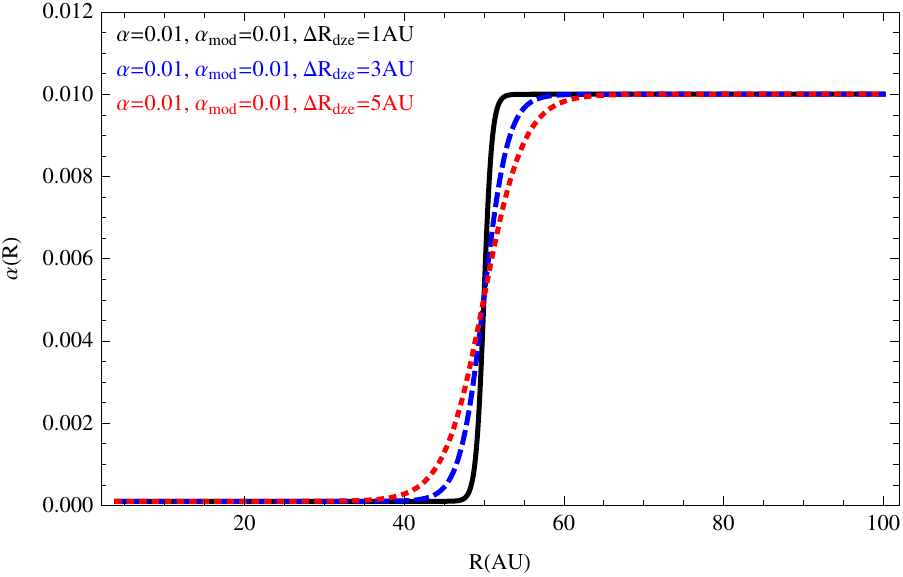}
	\caption{Example of smooth viscosity reduction according to Eq. (\ref{eq:deltaalpha}) used in our models. $\delta_\alpha(R)$ for models $\alpha=0.01$, $\alpha_\mathrm{mod}=0.01$, $R_\mathrm{dze}=50\,\rm AU$, and $\Delta R_\mathrm{dze}=1,3,5$\,AU are shown with \emph{black solid}, \emph{blue dashed}, and \emph{red dotted} curves, respectively.}
	\label{fig:alpha_damping}
\end{figure}

We revisit the development of RWI, \citep{Lovelaceetal1999}, by 2D, grid-based, global hydrodynamical disc simulations, assuming the $\alpha$-type viscosity prescription \citep{ShakuraSunyaev1973}. To model the dead zone, the value of $\alpha$, is smoothly reduced such that $\alpha(R)=\alpha\delta_\alpha(R)$ and 
\begin{equation}
	\label{eq:deltaalpha}
	\delta_\alpha(R)=1-0.5\left(1-\alpha_\mathrm{mod}\right)\left[-\tanh\left(\frac{R-R_\mathrm{dze}}{\Delta R_\mathrm{dze}}\right)+1\right],
\end{equation}
where $\alpha_\mathrm{mod}$ is the depth of turbulent viscosity reduction, $R_\mathrm{dze}$ and $\Delta R_\mathrm{dze}$ are the radial distance and the width of the transition region of turbulent viscosity at the dead zone outer edge, respectively. Regarding the present study the effect of the inner edge of the dead zone is not relevant, therefore it is not investigated here. Examples of $\alpha(R)$ used in our models are shown in Fig.\,\ref{fig:alpha_damping}.

To quantify $R_\mathrm{dze}$, we adopt the results of \citet{MatsumuraPudritz2005}. According to their calculations, the outer edge of the dead zone lies between 12\,AU  and 36\,AU, depending on their various model parameters. To induce the vortex formation at distances suggested by submillimetre images of \citep{Brownetal2009}, we set the outer edge of the dead zone to slightly larger distance ($R_\mathrm{dze}=50$\,AU). Note that the exact location of the outer edge of the dead zone is very uncertain, because the physics of the ionisation (depending on the ionisation rate of the stellar X-rays, interstellar cosmic-rays, proton radiation from the high temperature disc corona, decay of radionuclides such as $\rm ^{40}K$ or $\rm ^{26}Al$) and the dependence of the magnetorotational instability on this ionisation degree is not yet well understood (see, e.g., \citealp{TurnerDrake2009}).

\subsection{Numerical simulations}

To follow the evolution of the RWI triggered vortices on several $10^6\,\rm yr$ time-scale, we used {\small GFARGO}\footnote{http://fargo.in2p3.fr/spip.php?rubrique18}, the GPU\footnote{Graphical Processor Unit constructed with several hundred procesors (240 for GTX-280, which we used for the simulations).} version of the {\small FARGO} code \citep{Masset2000}. We slightly modified the CUDA\footnote{http://www.nvidia.com/object/what\_is\_cuda\_new.html} kernel of {\small GFARGO} code in order to handle the viscosity reduction according to Eq.\,(\ref{eq:deltaalpha}).

The dimensionless continuity and Navier-Stokes equations 
, describing the surface mass density evolution in an accretion disc, are numerically solved in two-dimensional cylindrical coordinate system. We use the locally isothermal approximation, in which the vertically integrated gas pressure is $p(R)=c_\mathrm{s}(R)^2\Sigma(R)$, where $c_\mathrm{s}(R)$ is the local sound speed, and $\Sigma(R)$ is the vertically integrated surface mass density of gas. The initial surface mass density profile follows the power-law distribution, $\Sigma(R)\sim R^{-0.5}$. We use radially constant disc aspect ratio $h=H(R)/R$, where $H(R)$ is the pressure scale height at the radial distance $R$, for which case $c_\mathrm{s}(R)=\Omega H = \sqrt{G M_*/R^3}(h R)\propto R^{-1/2}$. 

\begin{figure}
	\centering
	\includegraphics[width=\columnwidth]{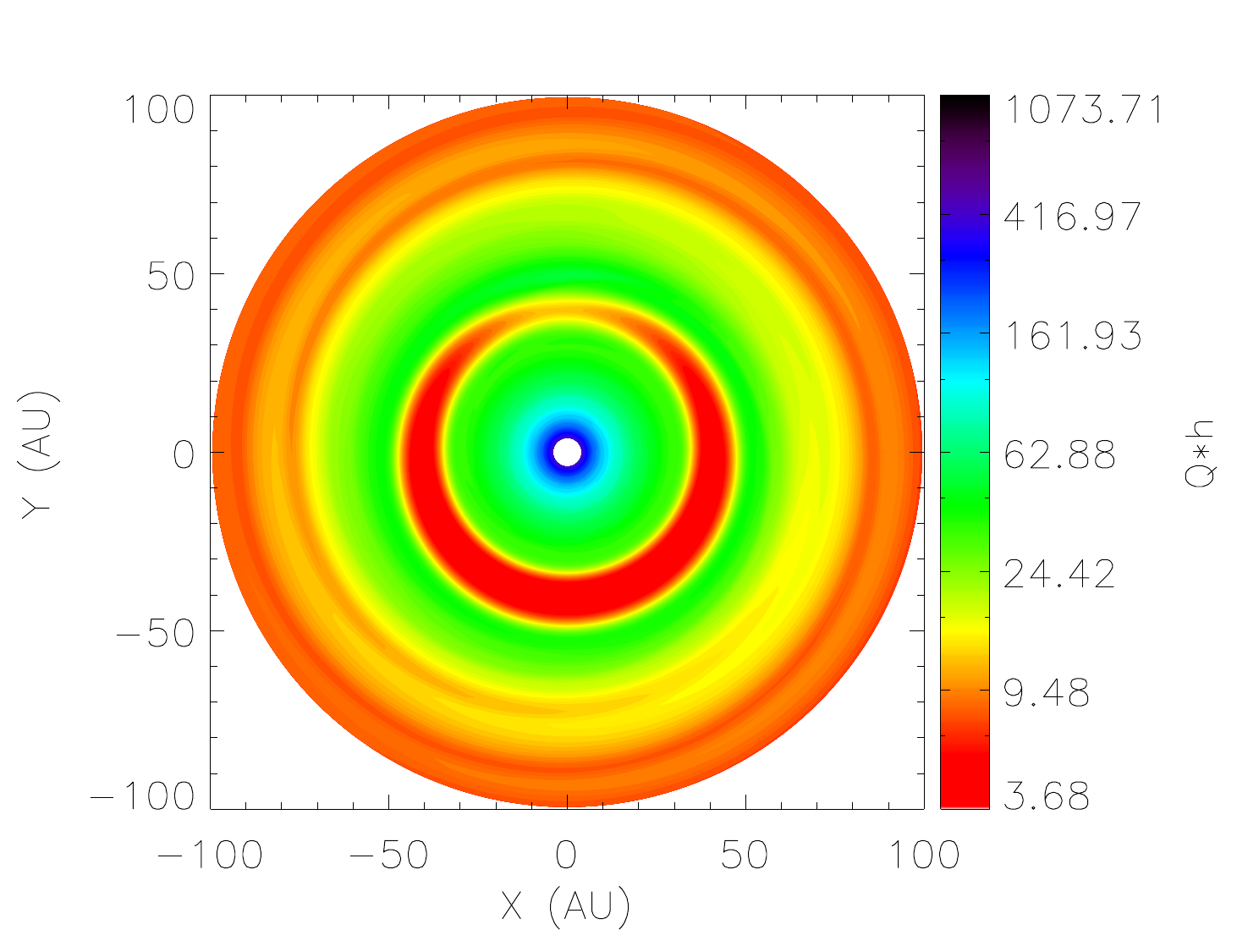}
	\caption{The distribution of $Qh$ (where $Q$ is the Toomre parameter and $h$ is the disc aspect ratio) calculated for the base model, when a large-scale vortex is fully-developed. Since $Qh$ is well above unity, even inside the vortex eye, seems reasonable to neglect the self-gravity.}
	\label{fig:Q2d}
\end{figure}

The disc self-gravity can be neglected as long as the Toomre parameter \citep{Toomre1964} is high, $Q=h M_*/(\pi R^2\Sigma)\gg1$, which holds in our unperturbed discs. Note, however, that $Q$ is relevant only for small scale, i.e. high $m$ modes, for global modes (e.g., m=1) the parameter that might really measure the relevance of self-gravity is $Qh$ \citep{Papaloizou2002}. Since $Qh$  might be smaller than unity, particularly inside the vortex owing to the density enhancement, we calculate $Qh$ for the base model ($h=0.05$ and $M_\mathrm{disc}=0.01\,M_\odot$, and neglecting self-gravity) by the time when the large-scale vortex have fully developed, see Fig.\,\ref{fig:Q2d}. As one can see, $Qh$ never reaches unity ($Qh|_\mathrm{min}=3.68$ inside the vortex eye), therefore we neglected the disc self-gravity in our models. Note that the disc mass $M_\mathrm{disc}=0.01\,M_\odot$ that we used in our simulations is in the range ($0.006\leq M_\mathrm{disc}\leq 0.026$) as has been derived by SED modelling for the three transition discs by \citet{Andrewsetal2011}. Nevertheless, $Qh$ might be smaller than unity for a disc which is three times more massive than the one we used, thus for that case one should take into account the self-gravity.

The disc inner and outer boundaries are fixed at 4\,AU\ and 100\,AU, respectively. The computational domain contains 256 radial logarithmically and 512 azimuthal uniformly distributed grid cells. The numerical convergence with the above resolution is found to be satisfactory, see Sect.\,\ref{sect:robust}. At the inner and outer boundaries the so called `damping boundary condition' are applied \citep{deValBorroetal2006mn}, where a solid wall is assumed with wave killing zones next to the boundaries.

\subsection{Development of a large-scale vortex}
\label{sec:vortex-devel}
\begin{figure}
	\centering
	\includegraphics[width=\columnwidth]{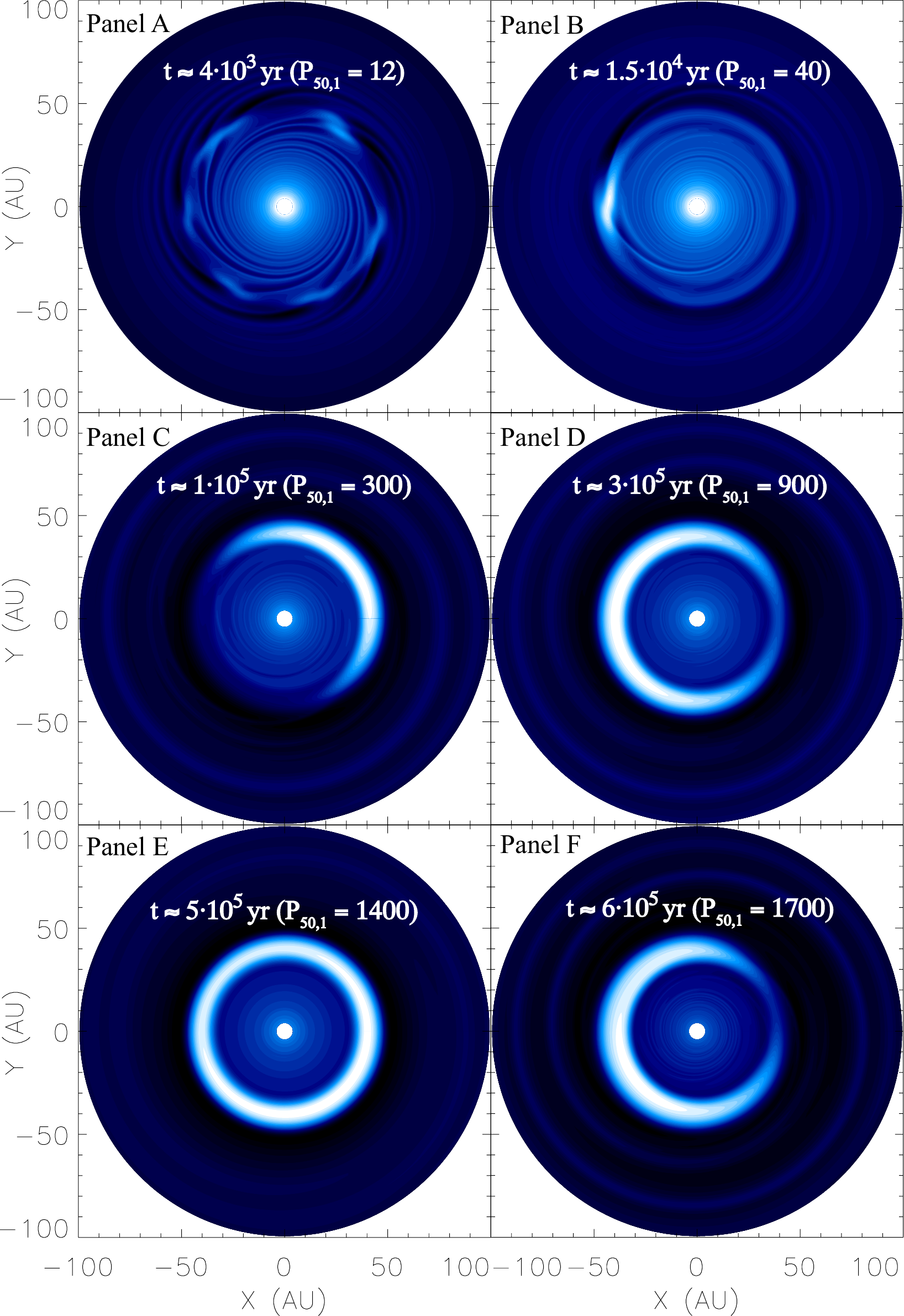}
	\caption{Major evolutionary stages of the surface mass density evolution in our base model ($M_{disc}=0.01\,M_\odot$, $H/R=0.05$, $\alpha=0.01$, $\Delta R_\mathrm{dze}=1\,\mathrm{AU}$, $\alpha_\mathrm{mod}=0.001$, ). \emph{Panel A:} excitation of the RWI at $\sim4\times10^3\,\rm yr$ ($P_{50,1}=12$). \emph{Panel B:} only one large-scale vortex remains after $\sim1.5\times10^4\,\rm yr$ ($P_{50,1}=40$). \emph{Panel C:} vortex evolution results in the formation of horseshoe-shaped (azimuthal wave number $m=1$ mode) surface mass density distribution at $\sim10^5\,\rm yr$ ($P_{50,1}=300$). \emph{Panel D:} vortex wings are interlocked at $\sim3\times10^5\,\rm yr$ ($P_{50,1}=900$). \emph{Panel E:} ring-like structure forms with small vortices on the top of that at $\sim5\times10^5\,\rm yr$ ($P_{50,1}=1400$). \emph{Panel F:} reappearance of the horseshoe-shaped vortex at $\sim6\times10^5\,\rm yr$ ($P_{50,1}=1700$).}
	\label{fig:vortex_evol}
\end{figure}

\begin{figure}
	\centering
	\includegraphics[width=\columnwidth]{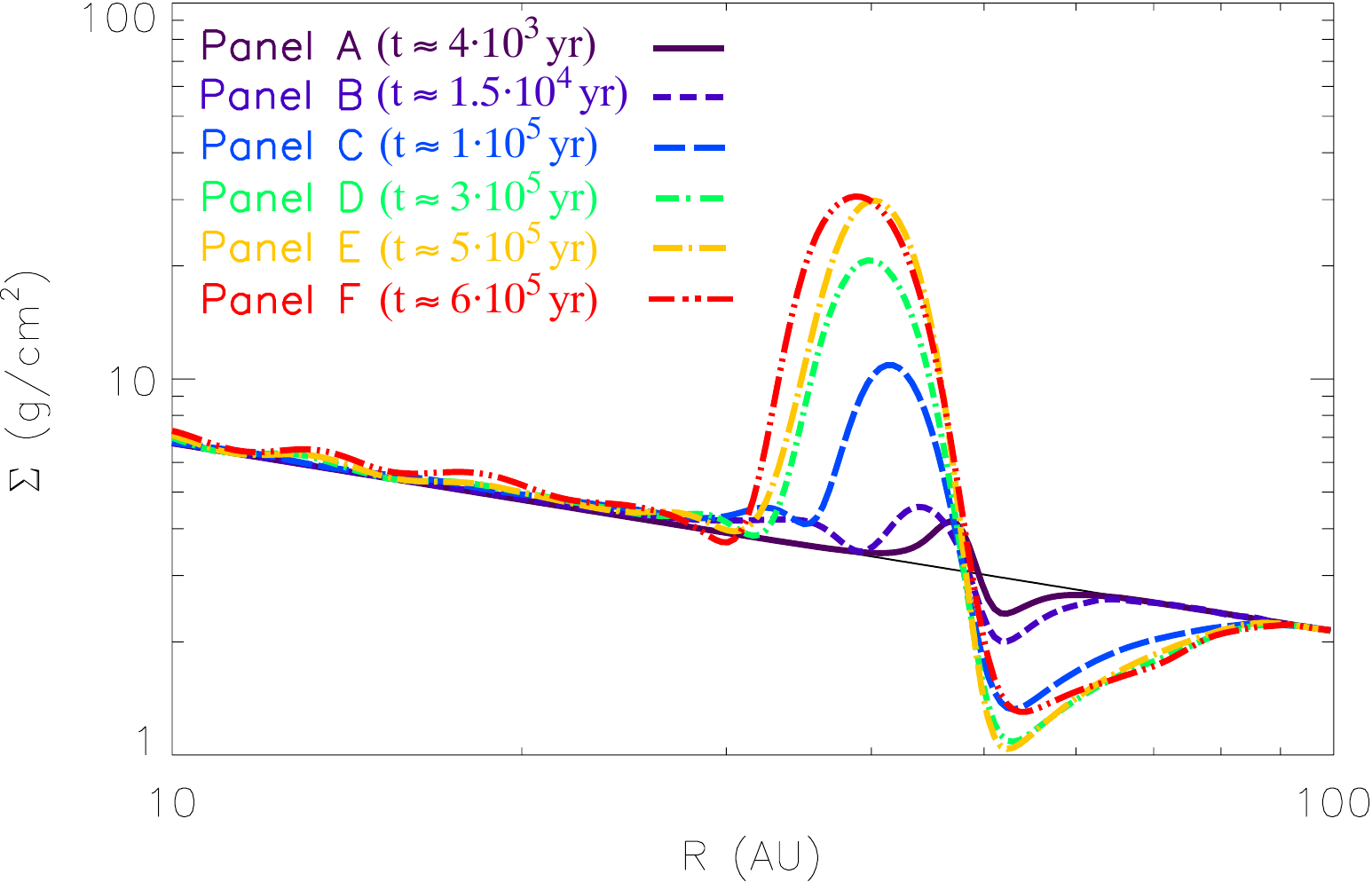}
	\caption{Evolution of the azimuthally averaged surface mass density profile in the base model. The \emph{solid black curve} represents the initial surface mass density profile. The density profiles shown with \emph{coloured dashed curves} are calculated from the surface mass density distribution presented in Fig.\ref{fig:vortex_evol}.}
	\label{fig:densprofile_evol}
\end{figure}

In what follows we give a short description of the formation and evolution of a large-scale vortex in our base disc model ($M_\mathrm{disc}=0.01\,M_\odot$, $h=0.05$, $\alpha=0.01$, $\alpha_\mathrm{mod}=0.001$, $R_\mathrm{dze}=\mathrm{50\,AU}$, and $\Delta R_\mathrm{dze}=1\,\mathrm{AU}$). The major evolutionary stages of the surface mass density distribution is shown in Fig.\ref{fig:vortex_evol}, while the azimuthally averaged surface mass density profiles are shown in Fig.\,\ref{fig:densprofile_evol}.

To excite RWI, a small mass orbiting body (negligible to the local disc mass) was put in the disc at 50\,AU. The orbiting perturber excites low-$m$ mode density waves emerging from its Lindblad resonances, and these in turn trigger the RWI. Triggering RWI by this method is appropriate, because RWI is a low-$m$ mode instability. We observed, though, that for models where the RWI is strong -- e.g., for sharp dead zone edge models ($\Delta R_\mathrm{dze}=1\,\rm AU$), where the radial gradients of pressure and density are steep -- the numerical noise in the radial velocity component can be large enough to trigger the vortex formation. 
This can be explained by the fact that the magnitude of radial velocity component is very small in Keplerian discs, thus a small numerical noise (which is in the order of $10^{-6}$ in the radial velocity component for our simulations done by {\small GFARGO}) is enough to excite RWI. For this reason, we investigated whether the mode of triggering RWI (by perturbing the radial velocity component or the surface mass density with superposition of sines or white noise; igniting density waves by a small mass planet orbiting at $\rm50\,AU$) has any effect on the vortex evolution. Studying carefully the results of our simulations, we have not found any significant differences, see details in Appendix\,B.


\begin{figure}
	\centering
	\includegraphics[width=\columnwidth]{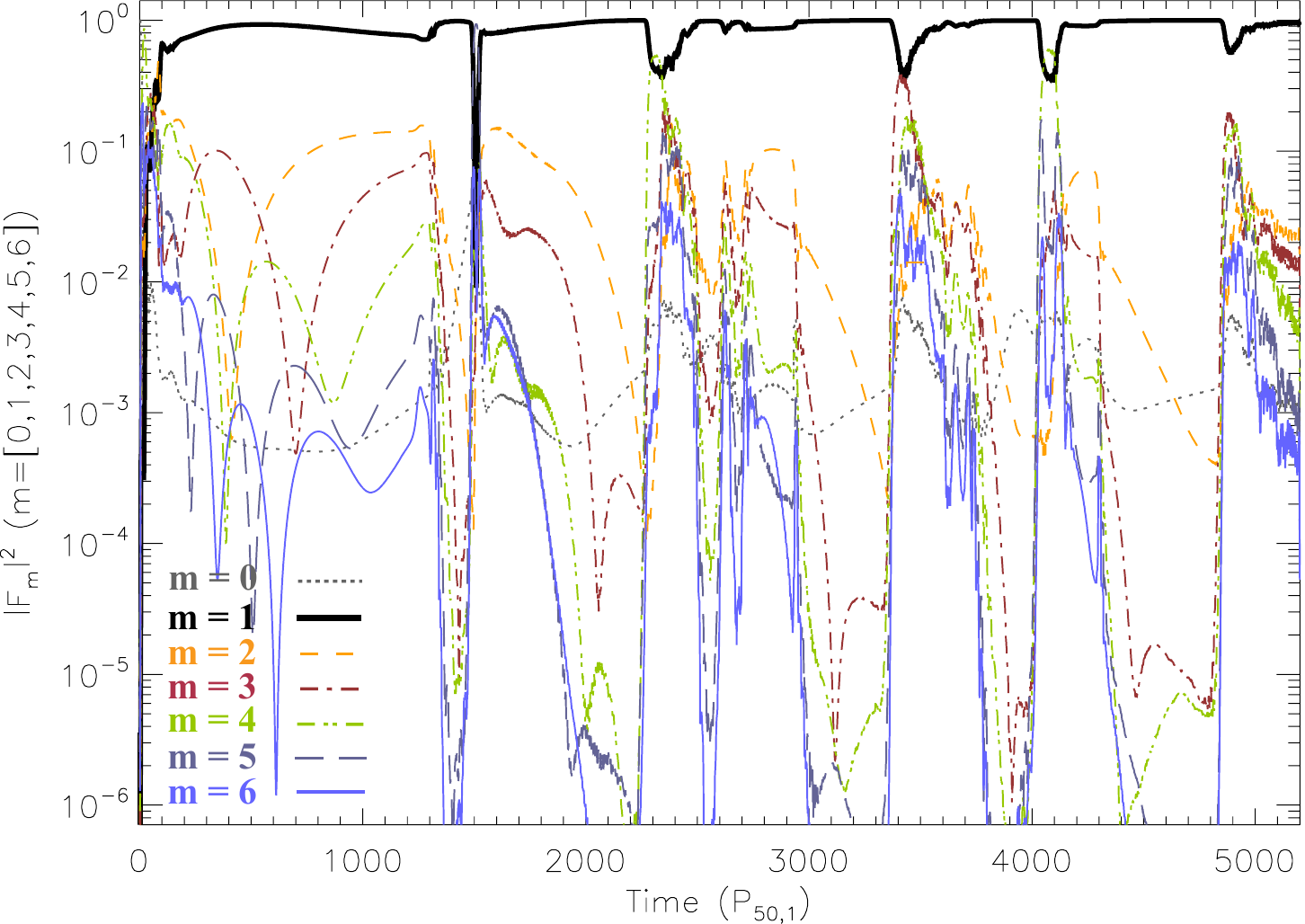}
	\caption{Azimuthal spectral power of the surface mass density distribution against time measured in $P_{50,1}$ for model presented in Fig\,\ref{fig:vortex_evol}. The Fourier amplitudes are normalised by the total power measured in the $m=0-6$ modes.}
	\label{fig:power_01}
\end{figure}

According to our simulation, the gas accumulates soon ($\leq10^2\,\rm yr$) at slightly smaller distance than the outer edge of the dead zone (see, e.g., Fig.\,\ref{fig:densprofile_evol} solid line curve). The RWI is excited at $\sim4\times10^3\,\rm yr$ ($12$ orbits at $50\,\rm AU$ assuming a Solar mass central star, i.e. $P_{50,1}=12$) forming several anticyclonic vortices with azimuthal wavenumber $m=6$ (Fig.\,\ref{fig:vortex_evol}, Panel A). These vortices coalesce, while the density bump continuously strengthens (see, e.g., Fig.\,\ref{fig:densprofile_evol} dashed line curves). As a result, only one large-scale vortex ($m=1$ mode vortex) remains by $\sim1.5\times10^4\,\rm yr$ ($P_{50,1}=40$) (Fig.\,\ref{fig:vortex_evol}, Panel B). Both the azimuthal and radial extensions of vortex increase, producing a horseshoe-shaped density perturbation by $\sim10^5\,\rm yr$ ($P_{50,1}=300$) (Fig.\ref{fig:vortex_evol} panel C). The vortex wings finally locks by $\sim3\times10^5\,\rm yr$ ($P_{50,1}=900$) (Fig.\,\ref{fig:vortex_evol}, Panel D). Investigating the following evolution of large-scale vortex in the base model, we found that the horseshoe-shaped density perturbation disappears, and a ring like density perturbation develops at $5\times10^5\,\rm yr$ ($P_{50,1}=1400$) (Fig.\,\ref{fig:vortex_evol}, Panel E). On the top of the ring-like density perturbation, small-scale vortices form, which are also subject to the vortex coalescing process (see in the next paragraph). After $\sim10^5\,\rm yr$ ($P_{50,1}=300$) of evolution, the $m=1$ mode (horseshoe-shaped) vortex reappears (Fig.\,\ref{fig:vortex_evol}, Panel F).

To illustrate this vortex mode oscillation we calculated the azimuthal spectral power of the surface mass density distribution against time measured in the orbital period $P_{50,1}$ shown in Fig\,\ref{fig:power_01}. First, a periodic Fourier transform of the density in the ring (period of $2\pi$ along $\phi$-direction at the radius where the vortices are) was calculated. Then the Fourier amplitudes of the $m=0-6$ modes are calculated normalised by the total power in these seven modes. It is clearly seen that  the spectral power is mainly in the $m=1$ mode (black curve) throughout the whole simulation. However,  by $5\times10^5\,\rm yr$ ($P_{50,1}=1400$) of disc evolution, the spectral power in $m=1$ mode temporarily drops. When the $m=1$ mode vortex decays, the spectral power of $m=4,5$ modes abruptly rises, i.e. RWI is re-excited. These newly formed vortices are also subject to coalescing and form again an $m=1$ mode vortex  by $\sim10^5\,\rm yr$ ($P_{50,1}=300$). Thus, the typical vortex mode is $m=1$ through $\sim 1.8\times10^6\,\rm yr$ ($P_{50,1}=5200$) of vortex evolution, which is interrupted by short-lived higher mode ($1<m<5$) vortex configurations. Note that the spectral power in $m=0$ characterising the average surface mass density drops off immediately as RWI is excited.

As shown in Fig.\,\ref{fig:densprofile_evol}, the density maximum is pushed inward from its original location, i.e., the $m=1$ vortex is slightly shrinked due to the extra pressure caused by the density build-up. As a result, the fully-developed $m=1$ mode vortex forms at slightly smaller distance ($R\simeq40\,\mathrm{AU}$) to the central star than the distance where the viscosity reduction ($R_\mathrm{dze}=50\,\mathrm{AU}$) begins. It is also appreciable that the density enhancement inside the vortex eye increases with time. Finally, emphasise that the surface mass density of the disc is significantly lowered next to the dead zone edge. This density rarefaction continuously strengthens with time, forming a deep hole in the disc by $\sim1.5\times10^6\,\rm yr$ ($P_{50,1}=4250$).

\subsection{Robustness of large-scale vortex development}
\label{sect:robust}

To examine the robustness (sensitivity to disc properties) of the development of the large-scale vortex, we performed several hydrodynamic simulations. By harnessing the power of the GPU, each simulation required only a day of computer time. For this parameter study the boundary conditions, the disc mass, and the disc aspect ratio were the same as in the previously shown base model. In what follows we summarise briefly the results of these simulations.

\subsubsection{Numerical resolution}

To check whether our numerical solution is satisfactory with the original grid ($N_R = 256, N_\phi=512$), we recalculated the base model with higher resolution, namely $N_R=512, N_\phi=1024$. We found that the results are practically identical, even by frame-to-frame comparison. Hence, the spatial resolution of the base model (256$\times$512) is sufficient for the following simulations.

\begin{figure}
	\centering
	\includegraphics[width=\columnwidth]{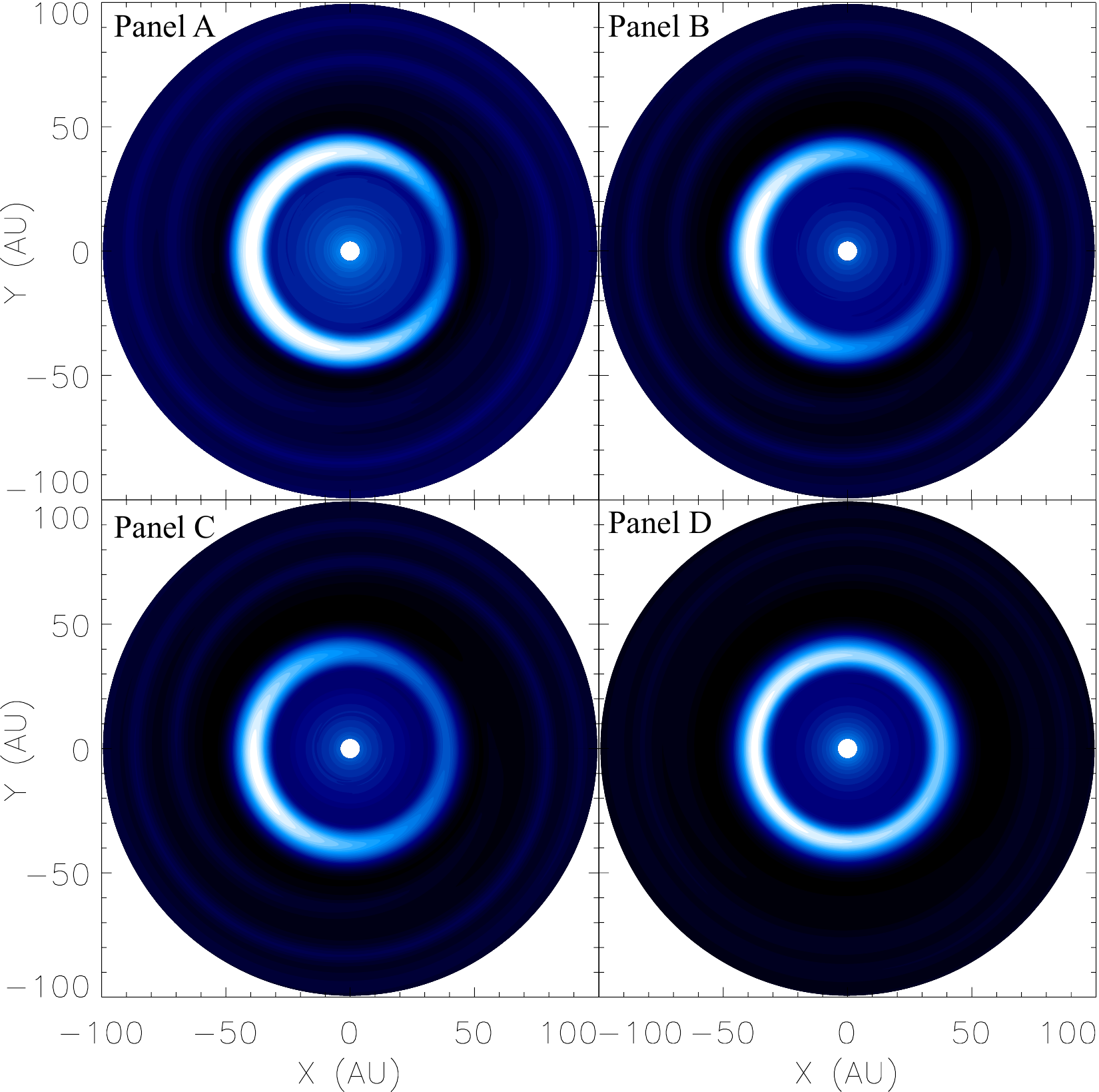}
	\caption{Fully-developed large-scale vortices at $3.5\times10^5\,\rm yr$ ($P_{50,1}=1000$) in models using $\Delta R_\mathrm{dze}=1,2$, and $3\,\mathrm{AU}$ dead zone edge width shown in panels A, B, and C, respectively. The axisymmetric $m=0$ configuration for model $\Delta R_\mathrm{dze}=5\,\mathrm{AU}$ shown in panel D is just the result of the viscosity reduction at the dead zone edge stalling the mass inflow, with subsequent pileup of gas.}
	\label{fig:vortex_dR1235}
\end{figure}

\begin{figure}
	\centering
	\includegraphics[width=\columnwidth]{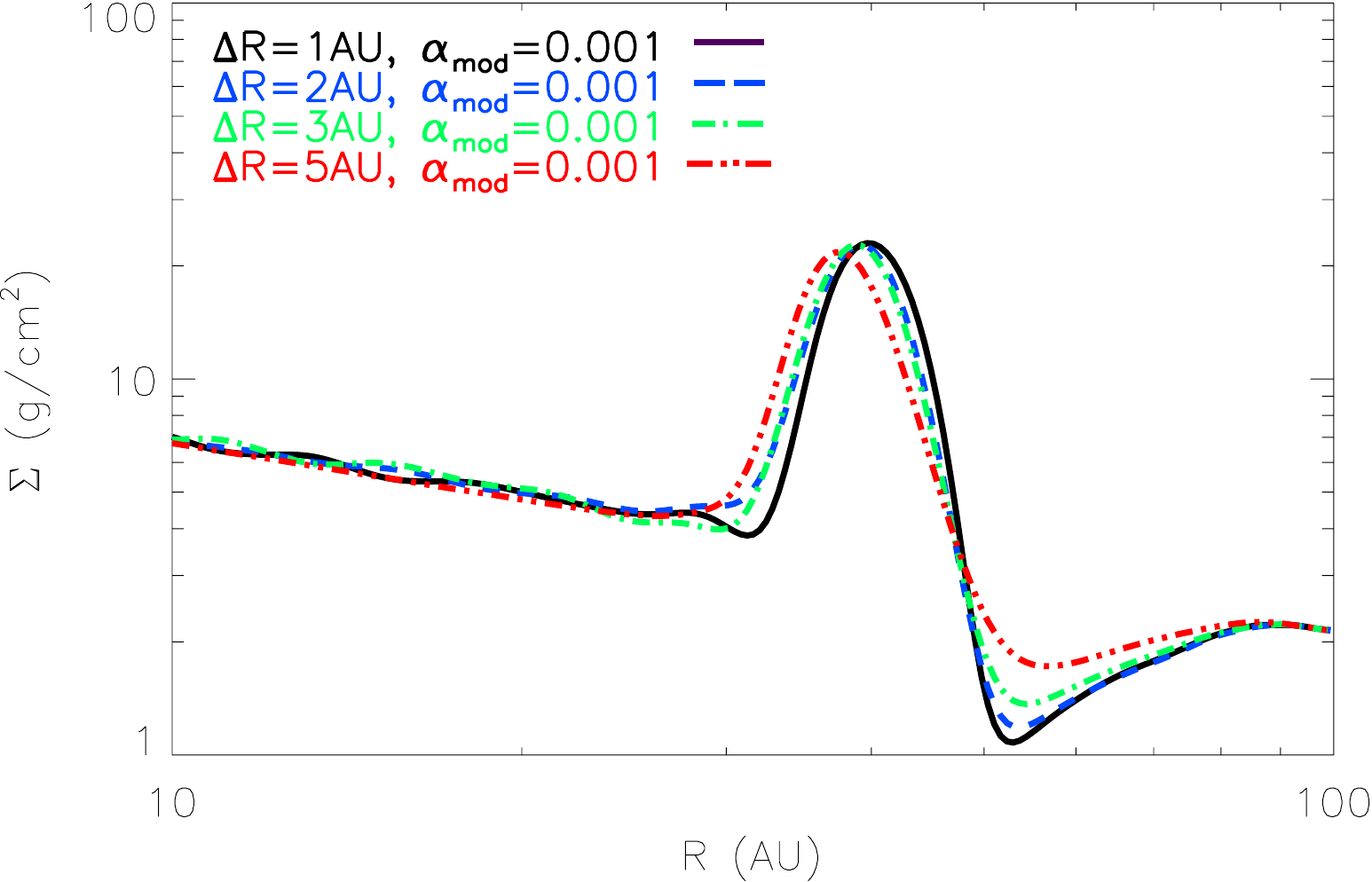}
	\includegraphics[width=\columnwidth]{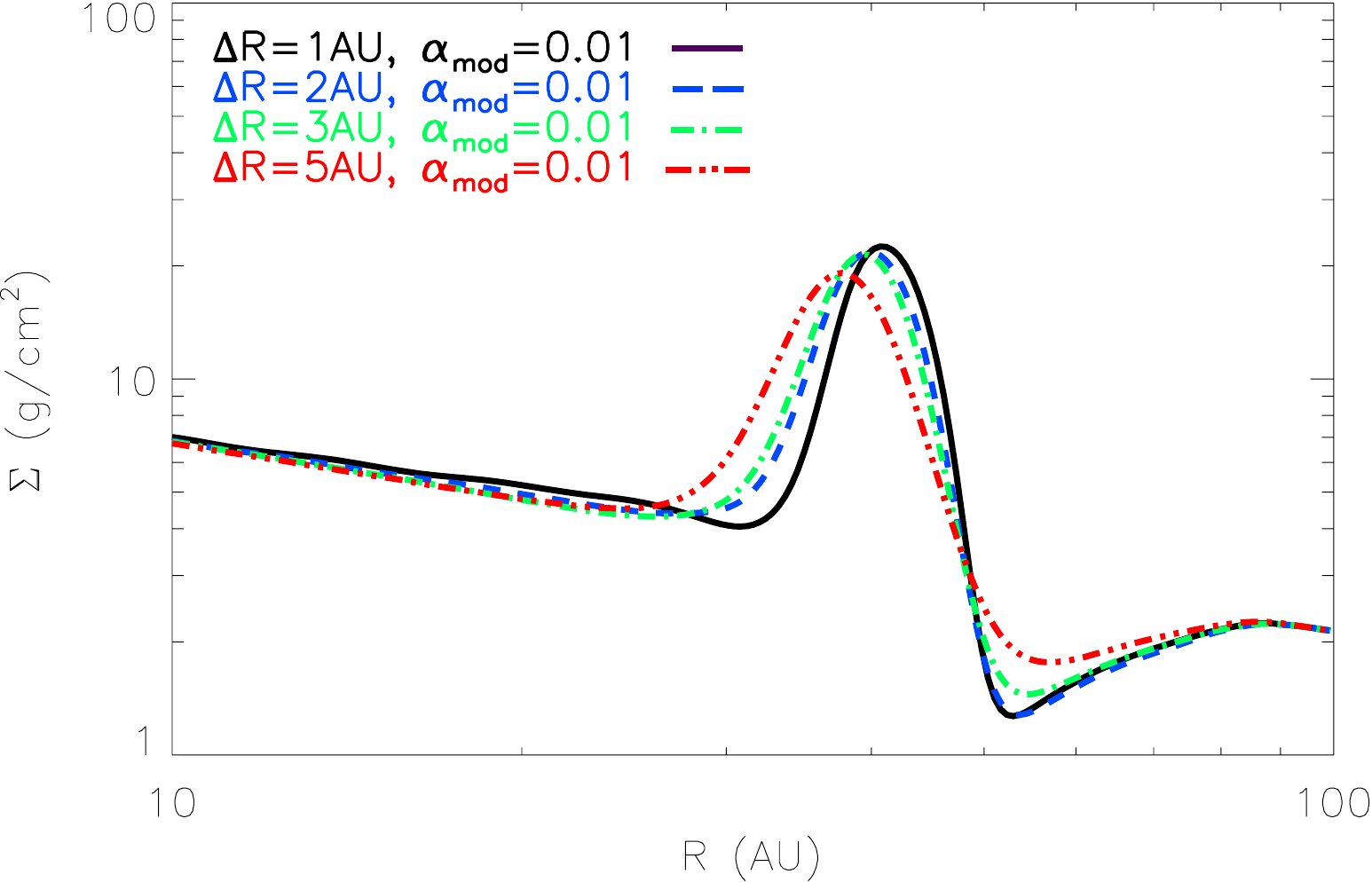}
	\caption{Azimuthally averaged surface mass density profiles at $3.5\times10^5\,\rm yr$ ($P_\mathrm{50,1}=1000$) in models using $\Delta R=1,2,3,\mathrm{and}\,5\,\mathrm{AU}$ dead zone edge width and $\alpha_\mathrm{mod}=0.001$ (\emph{upper panel}). Models with lower viscosity reduction $\alpha_\mathrm{mod}=0.01$ (\emph{lower panel}) show no significant departures to previous simulations.}
	\label{fig:densprofile_dRa}
\end{figure}

\begin{figure}
	\centering
	\includegraphics[width=\columnwidth]{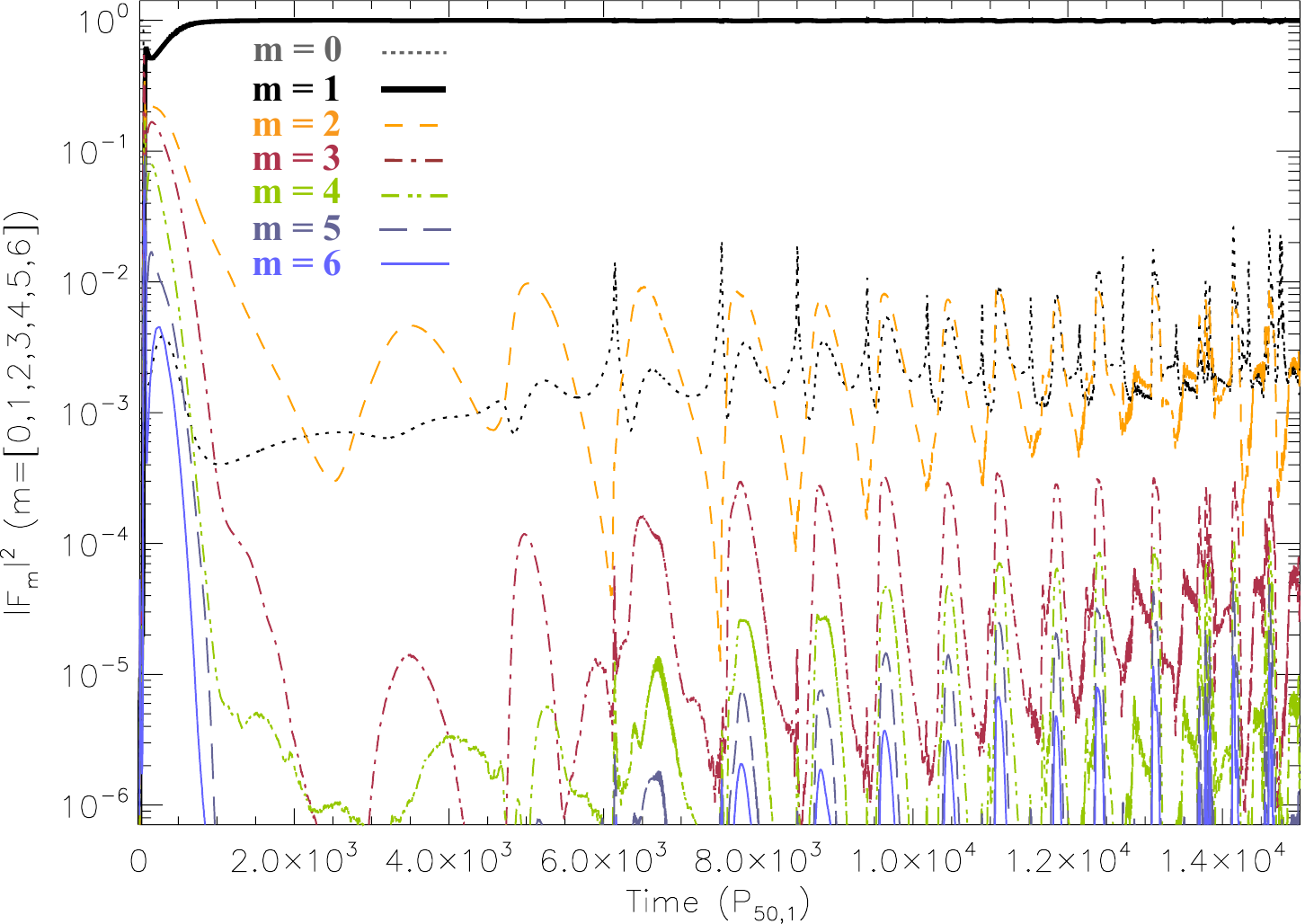}
	\caption{Azimuthal spectral power of the surface mass density distribution against time measured in $P_{50,1}$ for model $\alpha=0.01$, $\Delta R_\mathrm{dze}=3\rm \,AU$, $\delta\alpha=0.01$. The Fourirer amplitudes are normalised by the total power measured in the $m=0-6$ modes.}
	\label{fig:power_02}
\end{figure}

\subsubsection{Numerical precision}
Scientific calculations with conventional GPU cards (e.g., NVIDIA GTX-280 that we used) lacks the usual double precision at the moment.\footnote{Note that there exist scientific grade graphics cards in the NVIDIA Tesla series which are capable of double precision calculations.} To check the validity of our results provided by the single precision GPU we compare them to those of given by the CPU based {\small FARGO} code in the fastest evolving model ($\alpha=0.1$, $\Delta R_\mathrm{dze}=1\rm \,AU$, $\alpha_\mathrm{mod}=0.001$). The two runs give qualitatively the same result: the evolution of the large-scale vortex (growth rate, vortex coalescing, wing interlock, and oscillation) is found to be practically the same (see the details in Appendix A).

\subsubsection{Dead zone edge width}

\citet{Lietal2000} have found that the excitation of the RWI occurs later in time if the transition region of the viscosity reduction (i.e., the width of the dead zone edge) is wider. We confirmed this by setting the dead zone edge width $\Delta R_\mathrm{dze}$ between $1\,\mathrm{AU}-5\,\mathrm{AU}$ for models using $\alpha=0.01$ and $\alpha_\mathrm{mod}=0.01$. The excitation of the RWI occurs about $4$, $8.5$, and $20\times10^3\,\rm yr$ ($P_{50,1}=12,24$ and $57$) for models with $1$, $2$, and $3\,\rm AU$ dead zone edge width, respectively. However, in model with $\Delta R_\mathrm{dze}=5\,\rm AU$, the RWI is failed to be excited and only axisymmetric ($m=0$) density perturbation develops as a result of the viscosity reduction at the dead zone edge stalling the mass inflow, with subsequent pileup of gas. Thus, the RWI excitation occurs only in $\Delta R_\mathrm{dze}\leq2H$ models in a good agreement with \citet{Lyraetal2009}.

Figures\,\ref{fig:vortex_dR1235} and \ref{fig:densprofile_dRa} show the surface mass density contours and  the azimuthally averaged surface mass density profiles developed at $3.5\times10^5\,\rm yr$ ($P_\mathrm{50,1}=1000$) for various dead zone edge width models. Although the vortex structures are different (the larger the dead zone edge width, the smaller the azimuthal extension of the $m=1$ mode vortex, see Fig.\,\ref{fig:vortex_dR1235}, Panel A, B, C), the azimuthally averaged surface mass density profiles are very similar (Fig.\,\ref{fig:densprofile_dRa}, dashed line curves).

The vortex mode oscillation behaviour is completely absent in models with $\Delta R_\mathrm{dze}\geq2\,\mathrm{AU}$ during the whole simulation covering $\sim5\times10^6\,\rm yr$ ($P_{50,1}=14000$). Figure\,\ref{fig:power_02} show the azimuthal spectral power of the surface mass density distribution against time calculated as described before. For this case, the largest spectral power is in $m=1$ mode and remains nearly constant throughout the whole simulation after the initial vortex coalescing process.

\subsubsection{Depth of viscosity reduction}

We investigated the effect of the magnitude of $\alpha_\mathrm{mod}$ being in the range of $0.001-0.01$ for models with $\alpha=0.01$. We found that the time required to ignite the RWI, and to form the fully-developed large-scale vortex are independent of $\alpha_\mathrm{mod}$. Moreover, the magnitude of density contrast developed in the large-scale vortex is also independent of $\alpha_\mathrm{mod}$. As a consequence, the azimuthally averaged surface mass density profiles (Fig.\,\ref{fig:densprofile_dRa} calculated at $\sim3.5\times10^5\,\rm yr$ ($P_{50,1}=1000$) in weak viscosity reduction ($\alpha_\mathrm{mod}=0.01$, \emph{dashed line curves})) models are very similar to that of in models with stronger viscosity reduction ($\alpha_\mathrm{mod}=0.001$, \emph{solid line curves})). The life-time of the $m=1$ vortex, i.e., the vortex mode oscillation behaviour, is also found to be independent of $\alpha_\mathrm{mod}$: for sharp dead zone edge models the vortex mode oscillation do occur, while for thick dead zone edge models the $m=1$ mode vortex is long-lived.

\subsubsection{Magnitude of global viscosity}

We found that the vortex evolution is sensitive to the magnitude of the global viscosity $\alpha$ being in the range of $0.001-0.1$. The time required to excite RWI and to form the fully-developed large-scale $m=1$ mode vortex decreases with increasing $\alpha$. Both the radial extension of the large-scale vortex and the density contrast in the vortex increase with $\alpha$. These can be interpreted by the accelerated vortex evolution in models with high global viscosity.

For high viscosity models ($\alpha=0.1$), the inner disc becomes ever denser during the later ($t\gg5\times10^5\,\rm yr$) epochs, and the density contrast drops between the large-scale vortex and the inner disc. In consequence, the $m=1$ mode vortex survives only up to $\sim10^6\,\rm yr$ in these models. For low viscosity discs ($\alpha=0.001$), the vortex evolution slows down considerably, thus the excitation of RWI occurs later and the life-time of $m=1$ mode vortex can reach $\sim5\times10^6\,\rm yr$. 

We found that for low viscosity discs the vortex mode oscillation is completely missing, even for sharp dead zone edge ($\Delta R_\mathrm{dze}=1\,\rm AU$) models. Contrary to this, for high viscosity discs, the vortex mode oscillation does occur even for thick dead zones ($\Delta R_\mathrm{dze}<5\,\rm AU$) models. Emphasise that the $m=1$ mode (horseshoe-shaped) vortex might survive up to the life-time of the disc for $\alpha=0.001$, independently of the dead zone edge width being smaller than the twice of the local disc height ($\Delta R_\mathrm{dze}\leq5\,\rm AU$).

\section{Submillimetre images}

In order to show the resemblance of the observations of \citet{Brownetal2009} and signatures of large-scale Rossby-vortices in protoplanetary discs, we calculated submillimetre images using the 3D radiative transfer code RADMC-3D\footnote{http://www.ita.uni-heidelberg.de/$\sim$dullemond/software/radmc-3d}. First, the temperature structure of the disc was determined in a Monte Carlo simulation then the images at 340\,GHz (880\,$\mu$m ) were calculated via raytracing. Submillimetre Array (SMA) observations were simulated in the Common Astronomy Software Applications package (CASA).\footnote{http://casa.nrao.edu/index.shtml} 

\begin{figure}
	\centering
	\includegraphics[width=\columnwidth]{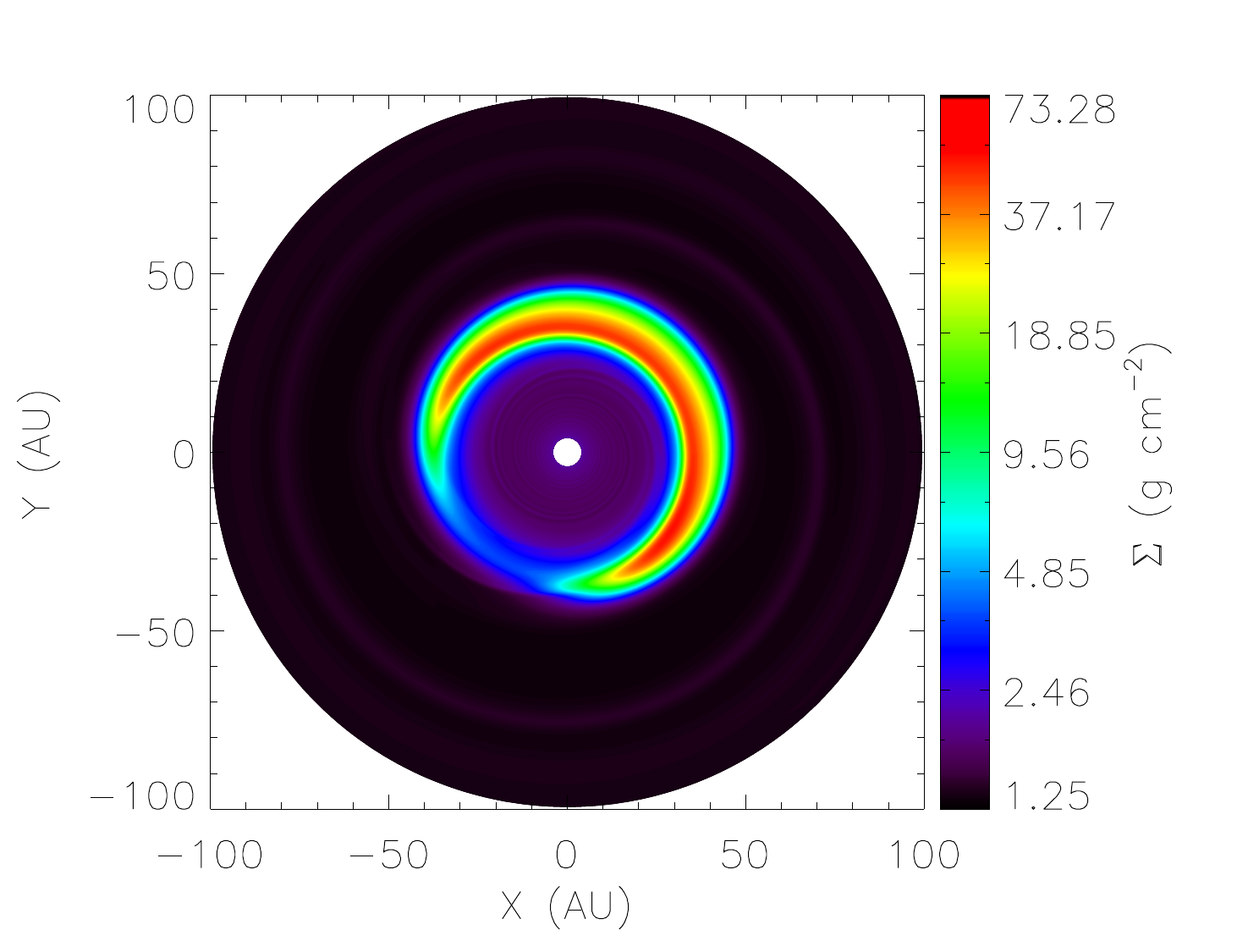}
	\includegraphics[width=\columnwidth]{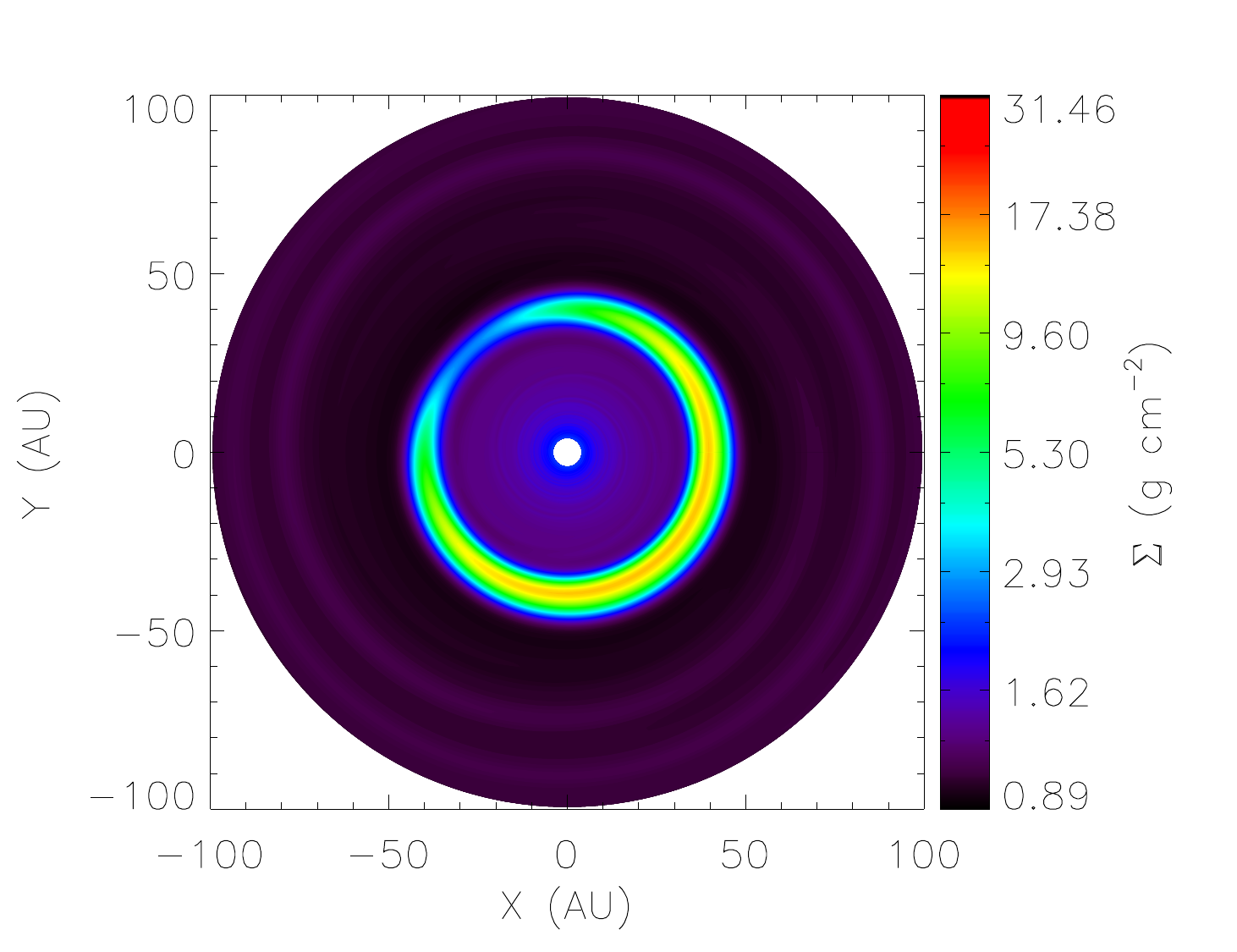}
	\caption{Hydrodynamic frames that were used to calculate the synthetic submillimetre images presented in Fig.\,\ref{fig:mm_images}. For modelling LkH$\alpha$\, 330 and SR\,21 frame shown in \emph{upper panel}, while for HD\,135344B frame shown in \emph{lower panel} were used. The frames were calculated in two different models: parameters were $R_\mathrm{dze}=50\,\mathrm{AU}$, $\Delta R_\mathrm{dze}=1\,\mathrm{AU}$, $\alpha=0.01$, $\alpha_\mathrm{mod}=0.001$ for \emph{upper panel}, and $R_\mathrm{dze}=50\,\mathrm{AU}$, $\Delta R_\mathrm{dze}=2\,\mathrm{AU}$, $\alpha=0.1$, $\alpha_\mathrm{mod}=0.01$) for \emph{lower panel}. To match the observed submillimetre images of \citet{Brownetal2009} frames had to rotate by $330^\circ$, $320^\circ$, and $170^\circ$ clockwise and spatially scale by 2, 1.2 and 1.7 for LkH$\alpha$\,330, SR\,21N, and HD\,135344B, respectively.}
	\label{fig:model_frame}
\end{figure}

Snapshots of surface mass density distributions shown in Fig.\,\ref{fig:model_frame} from the hydrodynamic simulations were converted into volume density assuming a vertically Gaussian density distribution with the same constant aspect ratio ($h=0.05$) as used in the hydrodynamic simulations. The geometrical dimensions of the sources as well as their distances were adapted from \citet{Andrewsetal2011}. The hydrodynamical frames had to be rotated and spatially scaled to approximately match the disc shapes and sizes on the measured submillimetre images. The used rotation angles were $330^\circ$, $320^\circ$, and $170^\circ$ clockwise, while the linear scaling factors were 2, 1.2, and 1.7 for LkH$\alpha$\,330, SR\,21N, and HD\,135344B, respectively. Since there are evidences for a fairly large gap in all discs \citep{Brownetal2009}, we reduced the dust density within a certain radius according to \citet{Andrewsetal2011}. Inside the dead zone, dust grains may grow and drift rapidly inwards leaving a reduced density region, i.e. a gap, behind. In anticyclonic vortices, however, dust grains get trapped and cannot drift inwards, which would qualitatively explain why the outer radius of the gap is close to the location of the RWI. Note that there is also evidence for the presence of an inner disc in these sources \citep{Fernandezetal1995,vanderPlasetal2008,Pontoppidanetal2008,Salyketal2009}. Since these inner discs have an outer radius probably smaller than the inner radius of our hydrodynamic simulations (4\,AU) we neglected them.

In the radiative transfer simulations we used the hydrodynamic grid in the radial and azimuthal directions while  N$_{\theta}$=100 grid points were used in the $\theta$ direction. The polar grid was a uniformly distributed, but we split the $-\pi/2\leq\theta\leq +\pi/2$ domain into two regions in order to resolve the disc vertically, while keeping the simulation computationally feasible. We placed 80 grid points in the $|\theta|\leq 0.35$ domain and 20 grid points were placed in the $|\theta|\geq 0.35$ region. In the thermal Monte Carlo simulations $2\times10^7$ photons were used to determine the temperature structure. 

\begin{figure}
	\centering
	\includegraphics[width=9.2cm]{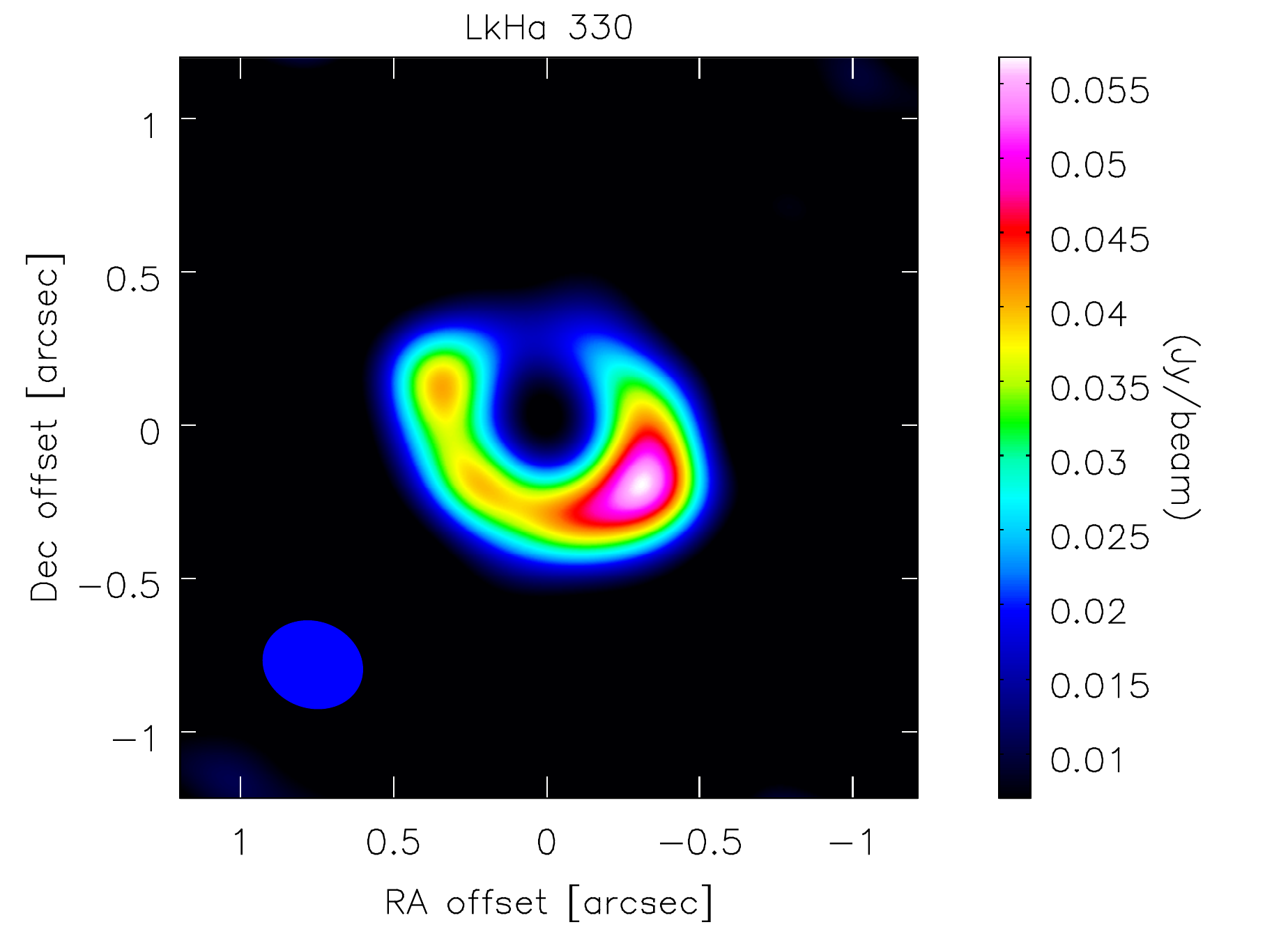}
	\includegraphics[width=9.2cm]{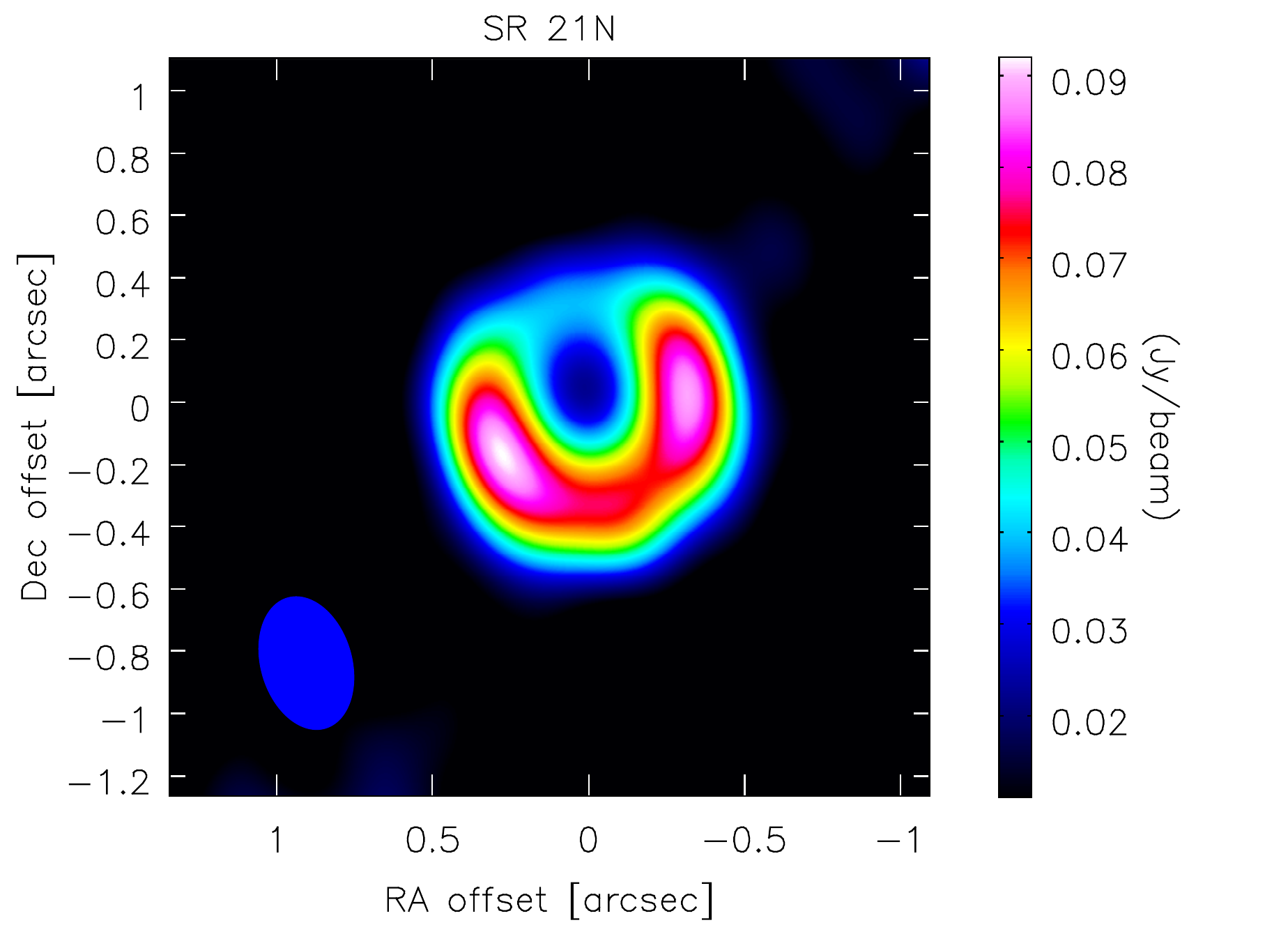}
	\includegraphics[width=9.2cm]{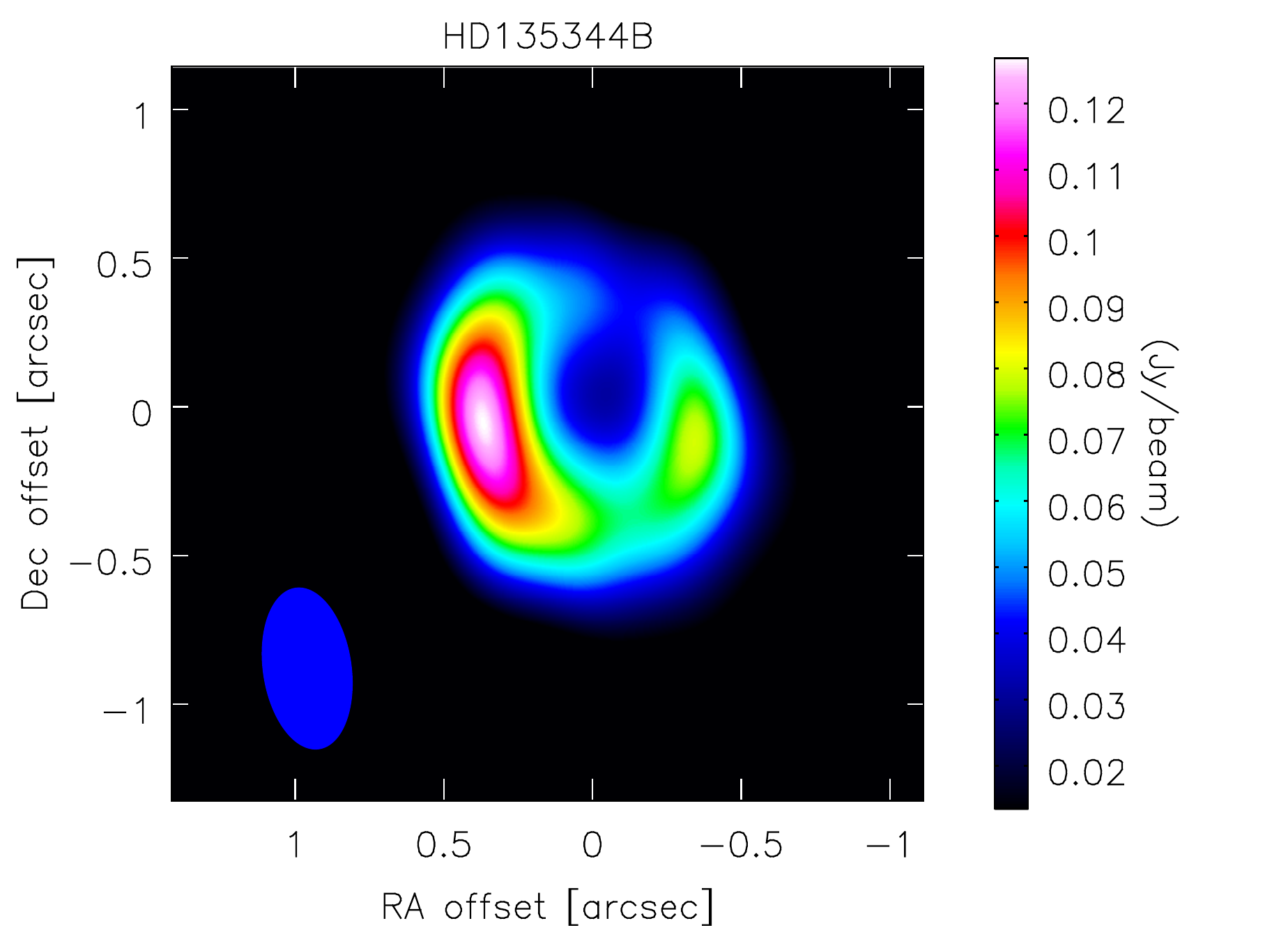}
	\caption{Submillimetre images of LkH$\alpha$\,330 (\emph{top}), SR\,21N (\emph{middle}) and HD\,135344B (\emph{bottom}) calculated by the 3D radiative transfer code, from the frames of the hydrodynamical calculations presented in Fig.\,\ref{fig:model_frame}. The synthesised beam is shown in the lower left corners in blue. The position angle of the discs, i.e. the vortex eye position was rotated corresponding to the observation of \citet{Brownetal2009}.}
	\label{fig:mm_images}
\end{figure}

\begin{figure*}
	\centering
	\includegraphics[width=5.8cm]{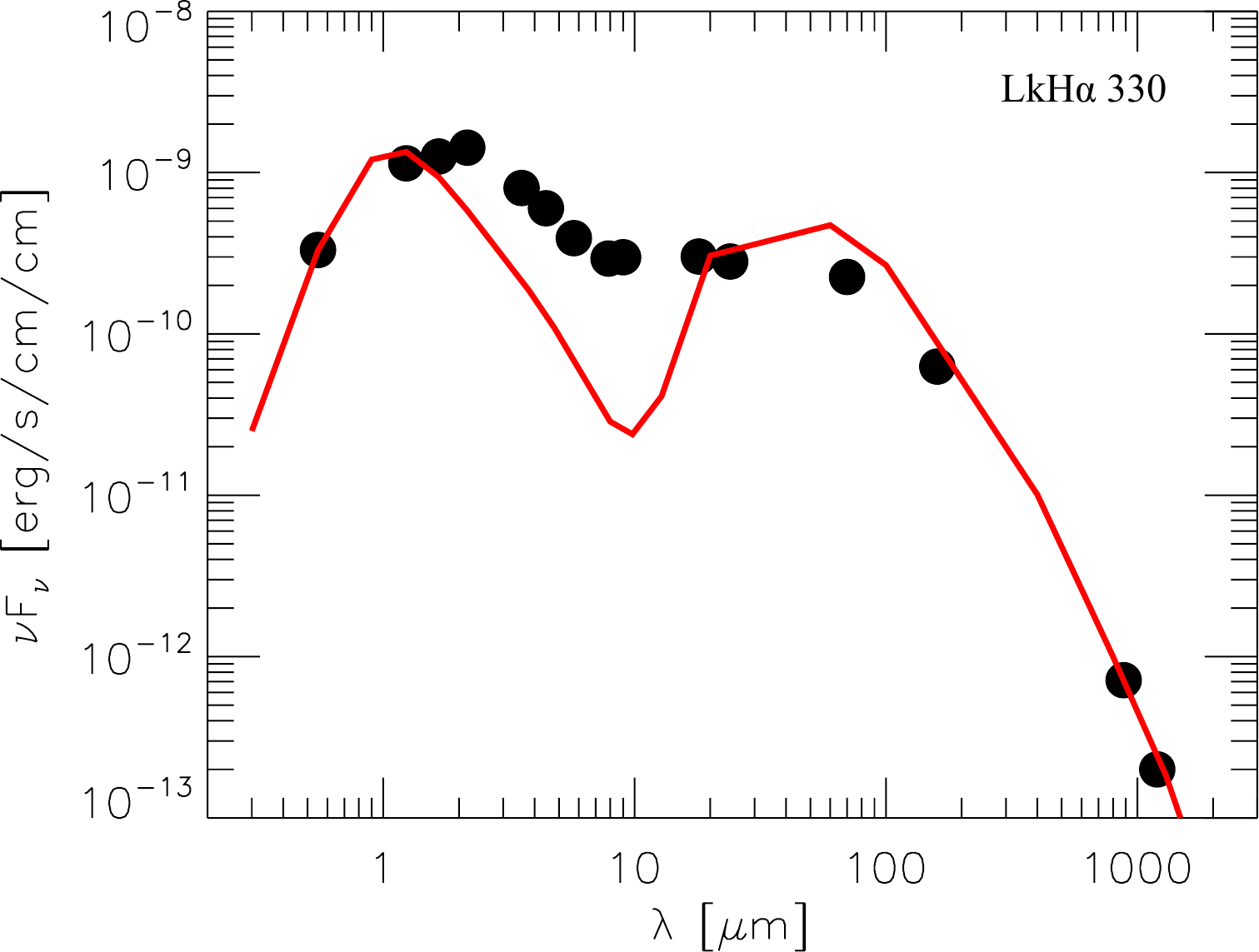}	
	\includegraphics[width=5.8cm]{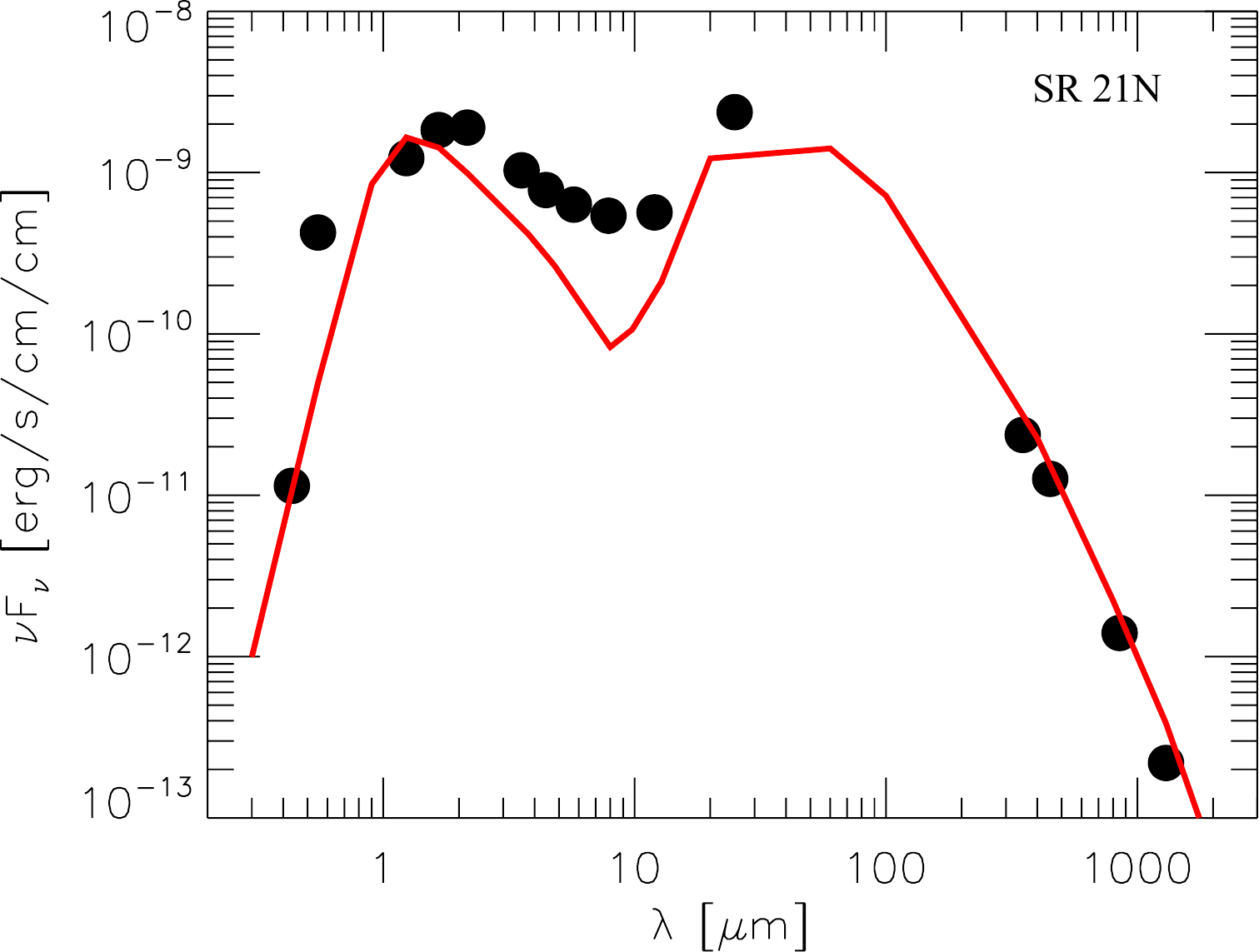}
	\includegraphics[width=5.8cm]{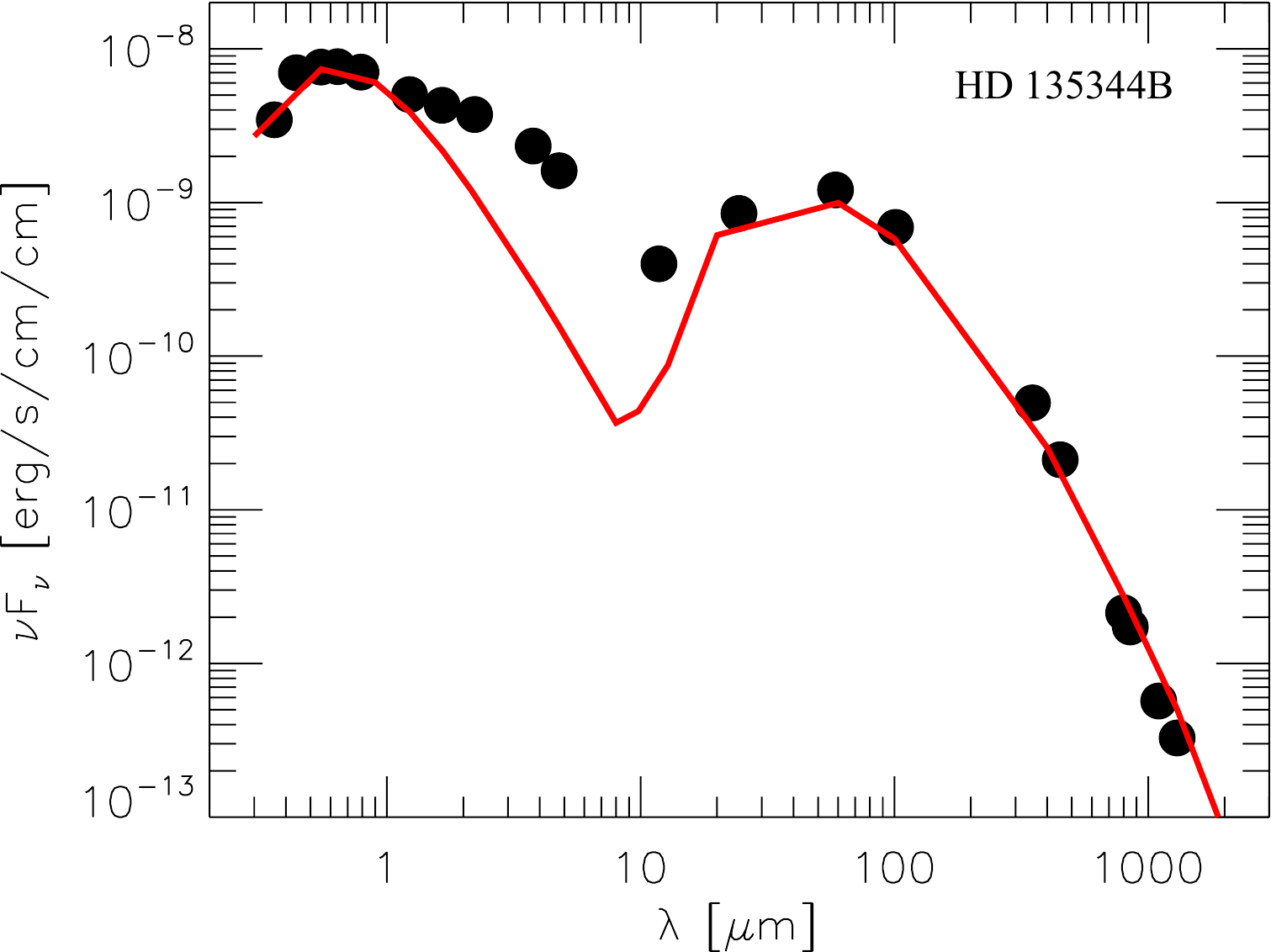}
	\caption{The observed (black dots) and the theoretical SEDs calculated in the ray-tracing models of LkH$\alpha$\,330 (\emph{left}), SR\,21N (\emph{middle}), and HD\,135344B (\emph{right}). The failure to fit the NIR flux is due to the fact that, for computational feasibility reasons, the inner disc emission is neglected.}
	\label{fig:SEDs}
\end{figure*}

The dust in our model consisted of ``astronomical'' silicate grains \citep{WeingartnerDraine2001}, whose absorption and scattering cross sections were calculated using Mie theory from the optical constants. Dust grains had an MRN size distribution \citep{MRN1977}, $n(a)\propto a^{-3.5}$, sampled with 10 logarithmic spaced bins with minimum and maximum grain size of 0.1\,$\mu$m and 1000\,$\mu$m, respectively. Dust particles with different sizes have different temperatures, thus they are not thermally coupled to each other. The dust-to-gas ratio was assumed to be 0.01 and constant within the disc outwards of the outer radius of the gap. Inside the gap the dust-to-gas ratio should be lower than 0.01 if the gap is formed due to the growth and inward drift of dust grains. Finally, the simulation of the observation (Fourier transformation, u-v sampling, cleaning) was done with CASA. The resulting images are shown in Fig.\,\ref{fig:mm_images}. We also calculated SED from our synthetic disc model for all three sources and compared them to the observations (see Fig.\,\ref{fig:SEDs}).

Our synthetic model images (Fig.\,\ref{fig:mm_images}) resemble the most important features of the submillimetre continuum map of the three transition discs presented in \citet{Brownetal2009}. The disc emission is clearly non-axisymmetric, horseshoe-shaped, and it shows single arc-like maximum at position angle around 225$^\circ$ for LkH$\alpha$\,330, 120$^\circ$ for HD\,135344B, while two maxima are visible at $120^\circ$ and $270^\circ$ position angles for SR\,21N. The SEDs (Fig.\,\ref{fig:SEDs}) calculated from our disc models match the observations shortwards of about 1\,$\mu$m and longwards of about 10\,$\mu$m. Between 1 and 10\,$\mu$m, however, our models under predict the observed fluxes, which is a natural consequence of neglecting the inner disc emission. We concluded that our disc models with large-scale Rossby vortices fit the observations of the mid-infrared to submillimetre emitting regions of all three discs.

Given the fact that our model is vertically optically thin at submillimetre wavelengths the observed asymmetry in the continuum emission can arise from temperature, density perturbations or the combination of both. Since our disc is optically thin at 880\,$\mu$m the increase in the column density will increase the observed flux. The higher density regions, however, can be shielded from the heating radiation (higher optical depth) and therefore the temperature can decrease. The temperature perturbations in our model are less than a factor of two, which means the same factor for the flux perturbation caused by temperature effects, since we are in the Rayleigh-Jeans regime. The density perturbations on the other hand are much larger in amplitude (factor of 6--8, see, e.g., the red dashed curve in Fig.\,\ref{fig:vortex_evol}) and are therefore most likely responsible for the observed asymmetries in the image. A detailed study of these effects in non-axisymmetric discs will be the topic of a forthcoming paper.

\section{Discussion \& Conclusion}

In this Paper we propose an interesting physical phenomenon that can be responsible for the large non-axisymmetric density perturbations revealed by submillimetre images of three transition discs of \citet{Brownetal2009}. Based on the results of our long-term, two-dimensional, grid-based, global hydrodynamic simulations (using compressible-gas model, neglecting the thermodynamical effects, and the disc self-gravity), we suggest that the observed asymmetries are caused by a large-scale vortex developed where the turbulent viscosity has a steep gradient, e.g. at the outer edge of the disc dead zone. Vortices are triggered by the Rossby wave instability (RWI), which later are subject to a coalescing process and form a large-scale anticyclonic vortex. Note that \citet{WolfKlahr2002} have previously shown that such large-scale anticyclonic vortices formed in the inner disc ($<10\,\rm AU$) can causes detectable intensity asymmetry in millimetre wavelengths ($345\,\rm GHz$ and $900\,\rm GHz$) which are proposed to be observed by the fully operable Atacama Large Millimeter/submillimeter Array (ALMA) with the longest available baseline.

After the initial excitation of RWI, a large-scale anticyclonic vortex develops by the coalescing of small-scale vortices at $R\simeq50\,\mathrm{AU}$, near the outer edge of disc dead zone, (Fig.\,\ref{fig:vortex_evol} and Fig.\,\ref{fig:densprofile_evol}), which grows both in size and density contrast. The time required to form a $m=1$ mode large-scale vortex is found to be $\sim1.5-4\times 10^4\,\rm yr$ ($P_{50,1}=40-115$), depending on the dead zone edge width $\Delta R_\mathrm{dze}$ (the sharper the dead zone edge, the faster the vortex evolution) and the magnitude of global viscosity (the larger the viscosity, the faster the vortex evolution). The $m=1$ mode vortex temporarily disappears and reforms through RWI re-excitation and subsequent vortex coalescing processes (Fig.\,\ref{fig:power_01}) with a cadence of $\sim10^5\,\rm yr$ ($P_{50,1}=300$) for sharp dead zone edge models ($\Delta R_\mathrm{dze}<2\,\mathrm{AU}$) and large global viscosity ($\alpha \geq 0.01$). However, the $m=1$ mode vortex remains stabilised for $\sim5\times10^6\,\rm yr$ ($P_{50,1}=14000$) for thick dead zone edge models ($2\,\mathrm{AU}\leq\Delta R_\mathrm{dze}<5\,\mathrm{AU}$) or small global viscosity ($\alpha\leq0.01$). For even thicker dead zone edge models ($\Delta R_\mathrm{dze}\geq 5\,\mathrm{AU}$), $m=1$ mode vortex does not form, only $m=0$ density perturbation develops. In short, we conclude that the horseshoe-shaped vortex ($m=1$ mode) formed due to the viscosity jump near the disc dead zone edge being at $\sim 50\,\rm AU$ can survive up to the disc life-time ($t\simeq5\times10^6\,\rm yr$), if the turbulent viscosity is small ($\alpha=0.001$), and the width of the dead zone edge is smaller than twice of the local disc height ($\Delta R_\mathrm{dze}<2H$).

To demonstrate the observability of the large vortices, we calculated synthetic submillimetre images with a 3D radiative transfer code using the surface mass density distribution provided by our hydrodynamic simulations. The distribution of the dust, emitting in the submillimetre wavelengths, is assumed to follow that of the gas, owing to the dust collecting effect of the pressure bump. The submillimetre intensity distributions given by our ray-tracing models -- fitting the observed SEDs in the stellar and submillimetre emitting domain (Fig.\,\ref{fig:SEDs}) -- show significant azimuthal asymmetries (Fig.\,\ref{fig:mm_images}). Comparing our submillimetre predictions to the observations of three transition discs LkH$\alpha$333, SR\,21N, and HD\,135344B \citep{Brownetal2009}, we found astonishing similarities. Note that there exist further examples of discs showing non-axisymmetric submillimetre images, such as LkCa\,15 \citep{Pietuetal2006}, GM\,Aur \citep{Hughesetal2009}, SR\,24S \citep{Andrewsetal2010}, and J1604-2130 \citep{WilliamsCieza2011} for which the same phenomenon might be responsible. Since the formation process of planetesimals and planetary embryos could be accelerated by the dust collecting and embryo trapping effect of the anticyclonic vortices caused by the vortical motions and the pressure bump \citep{BargeSommeria1995, KlahrHenning1997}, the non-axisymmetric submillimetre images presented by \citet{Brownetal2009} might be interpreted as snapshots of active planet-forming regions.

Note that there are other possible routes to generate vortices in protoplanetary discs, which also might result in $m=1$ vortex formation. For the presence of gap-carving giant planets the gas is expelled from the planet orbit and settles at the gap edges resulting in density maximum which is RWI-unstable and develops into vortices as shown by \citet{deValBorroetal2007}. Another way to generate vortices is the baroclinic instability, proposed by \citep{KlahrBodenheimer2003}, and recently confirmed by several works (e.g., \citealp{Petersenetal2007a,Petersenetal2007b,LesurPapaloizou2010,LyraKlahr2011}).

As it was mentioned before, not only enhanced dust coagulation, fragmentation should also happen inside the vortices if the observed non-axisymmetric features of submillimetre images are indeed due to dust trapped inside them. Otherwise, the submilimetre intensity would be weak due to the decreasing dust opacity of large grains. Thus, this would constitute indirect evidence of elliptic instability \citep{LesurPapaloizou2009} inside vortices leading to 3D turbulence in the vortex core, which \citet{LyraKlahr2011} quantified in compressible simulations to be at the level of 10 percent of the sound speed.

Here we have to mention some of the caveats of our calculations. Two-dimensional and three-dimensional turbulence behave quite differently regarding the energy cascade, which is form small to large $m$ (direct cascader) in 3D but can be form large to small $m$ (inverse cascade) in 2D. 3D global disc simulations of \cite{Meheutetal2010} revealed that although the RWI results in significant vertical streaming, the growth rate and evolution of vortices in 3D is very similar to that of in 2D. This may indicate that the vorticity generation by the RWI is strong enough to counter the 3D vortex stretching (the mechanism responsible for the direct cascade), so that the vortices survive and cascade inversely via viscous merging.

Our simulations cover several million years of the disc evolution, yet, the disc photoevaporation acting on the same time-scale was neglected for simplicity. Since photoevaporation process disperse the gas content of the disc, this might cause inward migration of the dead zone edge. Thus, photoevaporation process might play an important role in vortex evolution, which will be addressed in a forthcoming paper.

\citet{LinPapaloizou2011} have investigated the role of self-gravity on vortex instabilities generated at the outer edge of the gap opened by a giant planet in locally isothermal disc models. They found that the unstable modes shifts to higher azimuthal wavenumbers assuming self-gravity: the wavenumber of the most unstable mode increases as Q decreases. For weak self-gravity ($M_\mathrm{disc}\leq0.024\,M_*$), a single vortex forms through coalescing, which process is delayed as $Q$ decreases. However, for higher disc mass ($M_\mathrm{disc}=0.031\,M_*$) the final configuration is a vortex pair, $m=2$ mode. Since we used $M_\mathrm{disc}=0.01\,M_\odot$, our models are in the weak self-gravity regime, where the coalescing process leads to the formation of $m=1$ vortex. 

Self-gravitating disc simulations require the inclusion of thermodynamics (the solution of energy equation) too \citep{Pickettetal1998,Pickettetal2000,Gammie2001}, otherwise the gas concentration would be over predicted. \citet{Lyraetal2009} found that assuming adiabatic equation of state for the gas, the pressure bump (that gives rise to RWI) sharpens considerably compared to their locally isothermal models, due to the high temperatures associated with compression. As a result, RWI might be triggered easier in adiabatic than in isothermal model. \citet{Petersenetal2007b} investigated the baroclinic feedback, in which a vortex enhances azimuthal temperature gradients to reinforce the vortex itself, in global disc simulations with anelastic-gas model. They concluded that this baroclinic feedback mechanism producing stronger local temperature gradients results in long-lived vortices if the radial temperature gradient is sufficiently large and either the radiative cooling time is small, or the thermal dissipation is high. On the other hand, vortices in the course of vortex coalescing may grow massive enough and reach the Jeans length, which provides a natural barrier to growth as shown by \citet{MamatsashviliRice2009}. In their shearing sheet simulations, assuming a simple cooling law with a constant cooling time and including self-gravity, the vortex starts to shock and radiate away the excess vorticity around the Jeans length. As a result, the vortices can not grow above the Jeans length and short-lived, if there is no continuous vorticity production like for RWI, for which case the shear is continuously being converted into vorticity \citep{Lyraetal2009}. In our simulations, however, the vorticity is not generated by the local advective heat transport enhancing the azimuthal temperature gradients (see, e.g., \citealp{Petersenetal2007b}), but the steep pressure bump and azimuthal density perturbations formed near the dead zone edge. In short, investigating the vortex evolution and longevity in a self-gravitating global disc model, considering non-isothermal and compressible-gas model, would be important, and will be addressed in a forthcoming paper.

Finally, worth noting that multiple viscosity jumps might exist in the disc, e.g., at the inner edge of the dead zone or at the snow line \citep{KretkeLin2007}, which could also result in the formation of large-scale vortices. In a non-magnetised disc its self-gravity causes outward angular momentum transport via quasi-steady self-gravitating turbulence, which results in the generation of effective viscosity \citep{Pringle1981}. Since only the outer disc ($R\ga$100\,AU) is expected to become gravitationally unstable in a standard protoplanetary discs, the self-gravity caused effective viscosity has a gradient there. In such systems the large-scale vortex formed beyond $\sim$100\,AU might result in vortex aided planet formation process at extremely high distances, such as observed around Fomalhaut \citep{Kalasetal2008mn}. As a conclusion, planets might be assembled in diverse fashion by these ``planetary factories'' located at different distances to their host star. In protoplanetary discs where the RWI is excited at smaller radii than we investigated, the development of large-scale vortices might lead to the formation of a planetary system similar to the Solar System.

\section*{Acknowledgments}

This research has been supported in part by DAAD-PPP mobility grant P-M\"OB/841/ and `Lend\"ulet' Young Researcher Program of the HAS. We are grateful to F. Masset who helped us making some necessary modification in his {\small GFARGO} code. We thank M.R. Hogerheijde and Ch. Brinch for the help and discussion on CASA data processing. We also thank to the referee Wladimir Lyra for thoughtful comments that helped to significantly improve the quality of the paper.

\bibliographystyle{mn2e}
\bibliography{mn-jour,regaly}

\appendix

\section{On the numerical precision}

\begin{figure}
	\centering
	\includegraphics[width=\columnwidth]{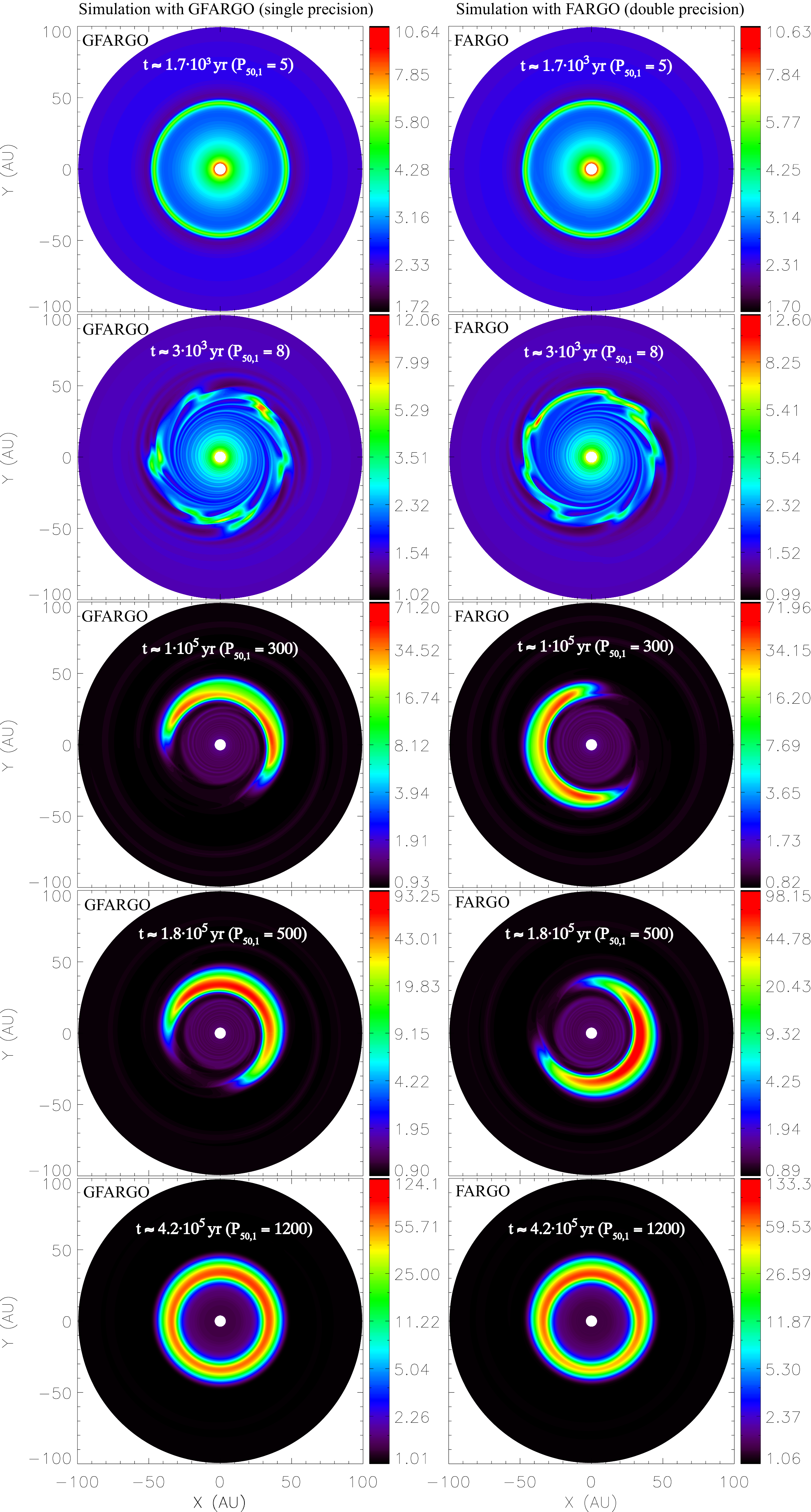}	
	\caption{Surface mass density evolution in a fast evolving model (i.e. model with $\alpha=0.1$, $\alpha_\mathrm{mod}=0.001$, $R_\mathrm{dze}=50\rm \,AU$, $\Delta R_\mathrm{dze}=1\rm \,AU$). The simulations were done by single precision {\small GFARGO} code (\emph{left column}) and double precision {\small FARGO} code (\emph{right column}) with exactly the same boundary conditions and numerical resolution. The surface mass density (shown with vertical colour bars for each frame) is measured in $g\,cm^{-2}$. The frames were saved at exactly the same time steps for both simulations. $P_{50,1}$ means the number of orbits at $50\,\rm AU$ assuming a Solar mass central star.}
	\label{fig:GFARGO-FARGO}
\end{figure}

As mentioned in Sec.\,\ref{sect:robust}, the GPU version of {\small FARGO} code ({\small GFARGO}) is only capable of single precision calculation at the moment, which obviously results in the degradation of precision. In order to be convinced that the results of {\small GFARGO} (single precision) are appropriate, we ran additional simulations by {\small FARGO} (double precision) for the model with $\alpha=0.1$ $\alpha_\mathrm{mod}=0.001$, $R_\mathrm{dze}=50\rm \,AU$, $\Delta R_\mathrm{dze}=1\rm \,AU$. In these simulations we applied the same boundary conditions, numerical resolutions ($N_r=256, N_{\phi}=512$), and disc masses for {\small GFARGO} and {\small FARGO}. The surface mass density distributions plotted at identical time steps for both simulations are shown in Fig.\,\ref{fig:GFARGO-FARGO}.

During both simulations, the RWI is excited with $m=6$ azimuthal wavenumber at $3\times 10^3\,\rm yr$ ($P_\mathrm{50,1}=8$). The subsequent vortex coalescing and vortex growth are qualitatively equivalent. However, the azimuthal angle of the $m=1$ mode vortices at $10^5\,\rm yr$ ($P_\mathrm{50,1}=300$) and $1.8\times 10^5\,\rm yr$ ($P_\mathrm{50,1}=500$) differs for the single and double precision simulations. This is because of the fact that $m=1$ mode vortex slowly drifts inward (see, e.g., Fig.\,\ref{fig:densprofile_evol}) due to the extra pressure caused by the density-enhancement at the dead zone edge, causing prograde precession of the vortex. Since the {\small FARGO} simulation results in slightly higher density peak than the {\small GFARGO} simulation (by $\sim5\%$ higher in average), the vortex drifts faster for the double precision, {\small FARGO} simulation. Consequently, the azimuthal angles of the vortices are different. 

\begin{figure}
	\centering
	\includegraphics[width=\columnwidth]{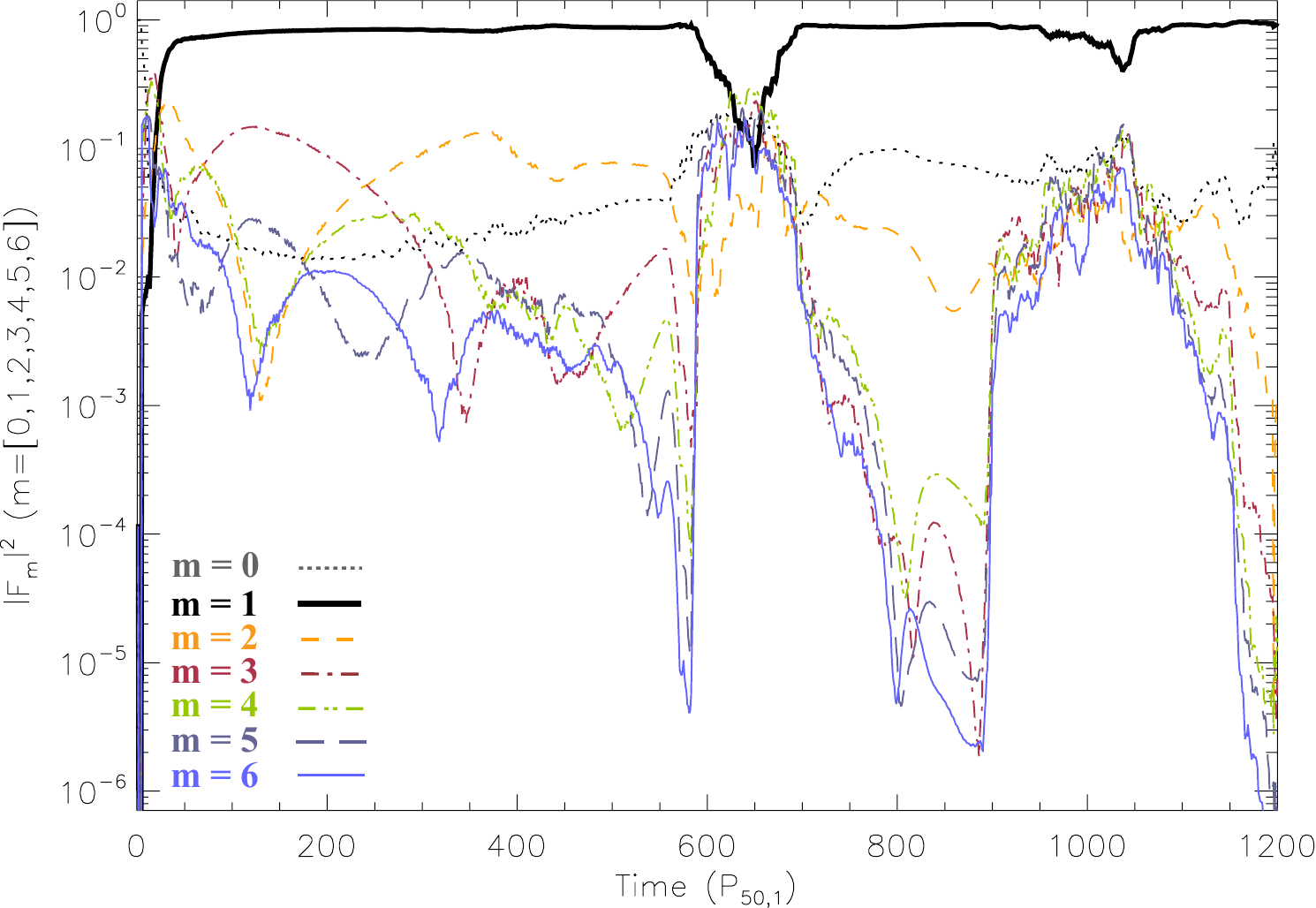}
	\includegraphics[width=\columnwidth]{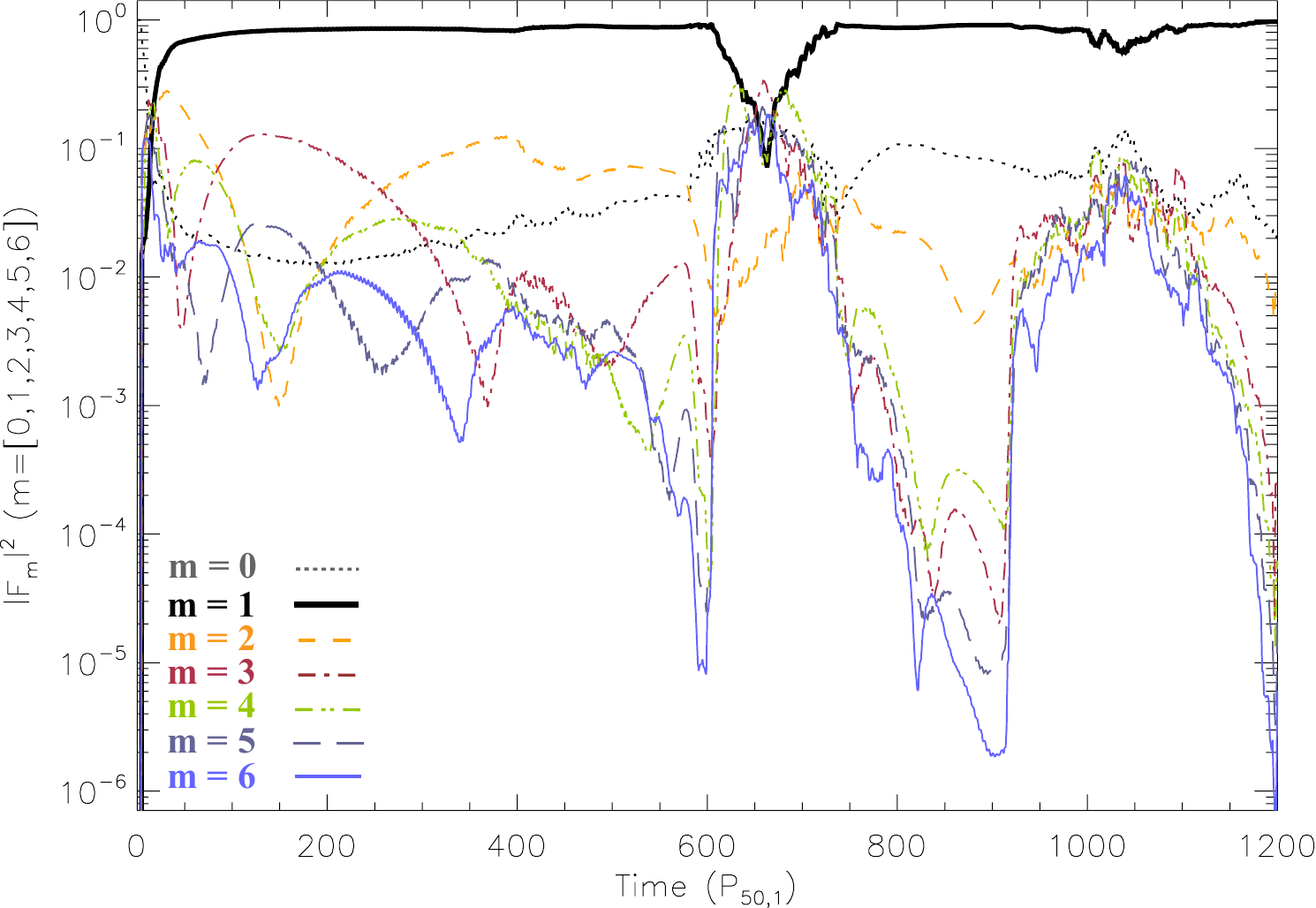}
	\caption{Azimuthal spectral power of the surface mass density distribution against time measured in $P_{50,1}$ for simulations shown in Fig.\,\ref{fig:GFARGO-FARGO}. Results given by {\small GFARGO} and {\small FARGO} are shown in the \emph{upper} and \emph{lower} panels, respectively. The Fourier amplitudes are normalised by the total power measured in the $m=0-6$ modes.}
	\label{fig:power_034}
\end{figure}

In order to investigate the vortex mode oscillation presented in Sect.\,\ref{sec:vortex-devel}, we calculated the azimuthal spectral power of the surface mass density distribution against time (Fig.\,\ref{fig:power_034}). As one can see, the evolution of the spectral power in the $0\leq m \leq 6$ modes are very similar for the two simulations. Note, however, that although the first and second vortex-decay occur at the same time for both simulations ($P_\mathrm{50,1}\simeq600$ and $P_\mathrm{50,1}\simeq1000$, respectively), the time required to reform the stable $m=1$ mode is slightly longer for the double precision simulation.

In short, we conclude that the lack of double precision capability of {\small GFARGO} is not dangerous on a million year time-scale in our simulations.

\section{On the excitation of RWI}

In order to make sure that no important feature is being missed during the excitation of the RWI, we ran several models where the RWI was seeded by conventional methods. To provide seeds for the RWI, we added low level perturbations to the surface mass density when the threshold for the RWI had been reached (i.e., when significant density-bump had been developed at the dead zone edge). In the first series of simulations, the perturbation was calculated as a superposition of sines, similarly to Eq. (10) presented in \citet{Meheutetal2010}:
\begin{eqnarray}
	\Sigma(R,\phi)^{'}&=&\Sigma(R,\phi) + |\Sigma(R,\phi)|\epsilon \sin \left( 2\pi R \right)\times \nonumber \\
	&&\left[sin(\phi) + \sin(2\phi) + \sin(3\phi) + sin(5\phi)\right],
	\label{eq:pert}
\end{eqnarray}
where $\Sigma(R,\phi)$, and $\Sigma(R,\phi)^{'}$ are the unperturbed, and perturbed surface mass density, respectively. The magnitude of the perturbation is set by $\epsilon$. In the second series, the surface mass density were perturbed by normal distribution white noise. While the former type of perturbation has a drawback that it excites only the low-$m$ modes as does a small mass orbiting planet, large part of the spectrum is being excited by the latter type of perturbation. We also tested the RWI excitation by small mass planet orbiting at $\rm 50\,AU$.

The RWI excitation was investigated in model where parameters of $\alpha=0.01$, $\alpha_\mathrm{mod}=0.01$, $R_\mathrm{dze}=50\,\rm AU$, and $\Delta R=1\,\rm AU$ were chosen, while the grid resolution was set to $N_R=256$ and $N_\mathrm{\Phi}=512$. The magnitude of perturbations were in the order of $\epsilon=10^{-4}$. The simulations were done by {\small FARGO}, because it has smaller numerical noise in the radial velocity component than in {\small GFARGO} simulations.

First, the threshold for the RWI was let to develop, i.e. a significant radial density gradient with magnitude of $\Delta\Sigma/\Sigma\sim 2$ at the dead zone edge. Then one of the above described perturbation was applied. To ascertain whether the RWI is not excited by the numerical noise of {\small FARGO}, we ran models without perturbation. In this unperturbed model, we found no RWI excitation.

\begin{figure}
	\centering
	\includegraphics[width=\columnwidth]{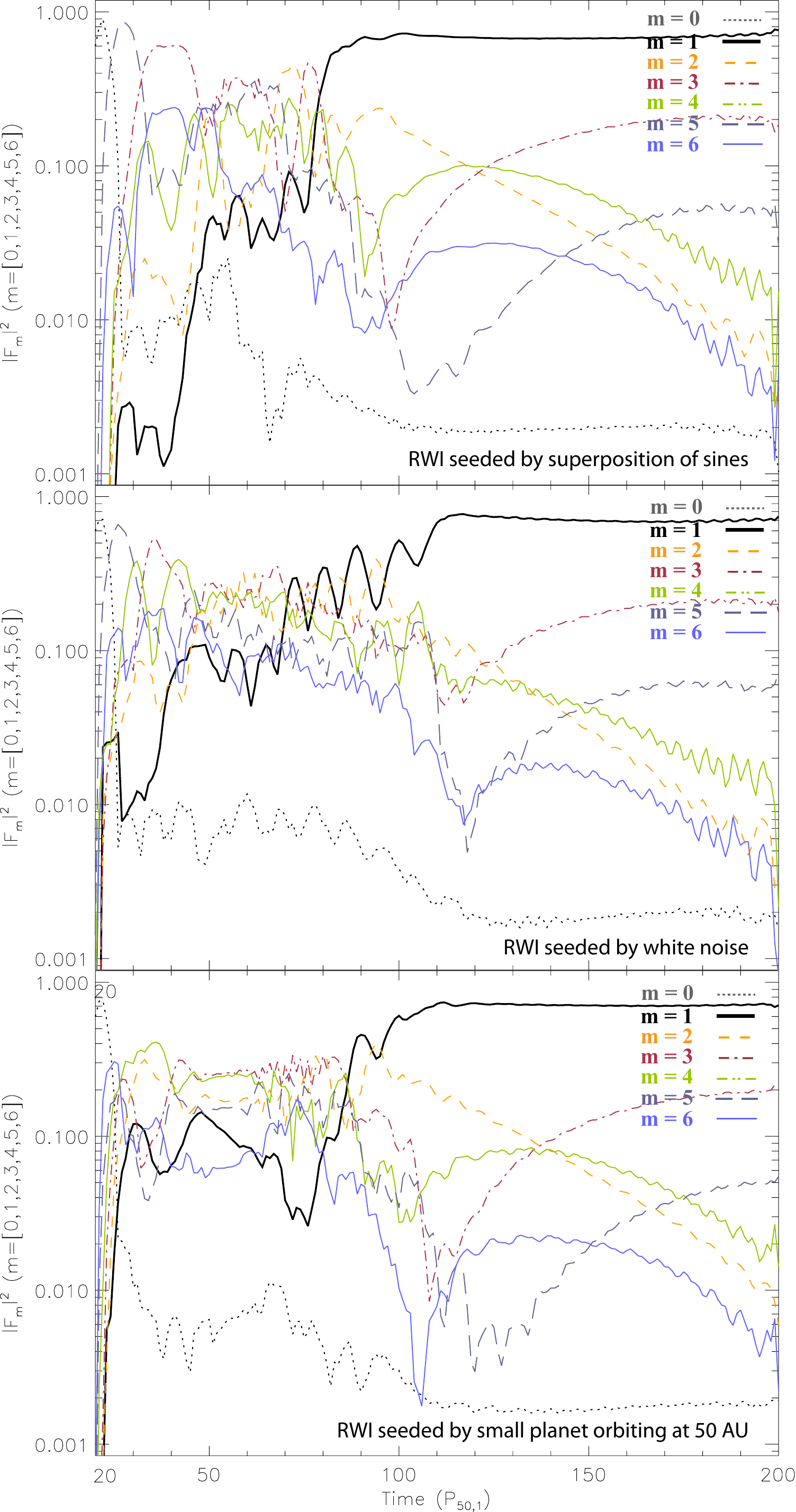}
	\caption{Evolution of the azimuthal spectral power of surface mass density distribution for models $\alpha=0.01$, $\alpha_\mathrm{mod}=0.01$, $\Delta R_\mathrm{dze}=1\rm \,AU$ followed as long as $m=1$ mode vortex had been fully fledged (at $P_{50,1}=200$). The RWI was seeded by superposition of sines (\emph{upper panel}), by normal distribution white noise (\emph{middle panel}) in surface mass density, and by small mass ($M_\mathrm{pl}=10^{-7}M_*$) planet orbiting at $\rm50\,AU$ (\emph{lower panel}). The Fourier amplitudes are normalised by the total power measured in the $m=0-6$ modes.}
	\label{fig:RWI-excitation_01}
\end{figure}

Figures \ref{fig:RWI-excitation_01} shows the evolution of azimuthal spectral power of the surface mass density followed as long as $m=1$ mode vortex had been fully fledged (by $P_{50,1}=150$) for models using different type of perturbations: superposition of sines in surface mass density with magnitude of $\epsilon=10^{-4}$ (\emph{upper panel}); normal distribution white noise in surface mass density with magnitude of $\epsilon=10^{-4}$ (\emph{middle panel}); small mass ($M_\mathrm{pl}=10^{-7}M_*$) planet orbiting at $\rm50\,AU$ (\emph{lower panel}). As one can see the evolutions of spectral power are very similar for any type of perturbations. Independent of the perturbation type applied, we found that i.) the RWI is always triggered after $P_\mathrm{50,1}\simeq5-10$ orbits at $\rm 50\,AU$; ii.) the most unstable mode of RWI is $m=5$ or $m=6$ at the onset of vortex formation; iii.) following the evolution of the vortices, $m=1$ mode (horseshoe-shaped) vortex always forms. Consequently, the mode of the excitation of RWI has no importance regarding our main findings that long-lived $m=1$ mode vortex appears near the disc dead zone edge.

\label{lastpage}
\end{document}